\documentclass{aa}
\usepackage{graphicx}
\usepackage{txfonts}
\usepackage[colorlinks,citecolor=blue,bookmarks=false,breaklinks=true]{hyperref}
\begin{document} 

\title{A$^{3}$COSMOS: Dissecting the gas content of star-forming galaxies across the main sequence at 1.2 $\leq z$ < 1.6}

\subtitle{}

\author{Tsan-Ming Wang
          \inst{1}
          \and
          Benjamin Magnelli
          \inst{2}
          \and
          Eva Schinnerer
          \inst{3}
          \and
          Daizhong Liu
          \inst{4}
          \and
          Eric Faustino Jim{\'e}nez-Andrade
          \inst{5}
          \and
          Christos Karoumpis
          \inst{1}
          \and
          Sylvia Adscheid
          \inst{1}
          \and
          Frank Bertoldi
          \inst{1}
          }

\institute{Argelander-Institut f\"ur Astronomie, Universit\"at Bonn, Auf dem H\"ugel 71, 53121 Bonn, Germany\\
              \email{twan@uni-bonn.de}
             \and
             Universit{\'e} Paris-Saclay, Universit{\'e} Paris Cit{\'e}, CEA, CNRS, AIM, 91191, Gif-sur-Yvette, France
             \and
               Max-Planck-Institut f\"ur Astronomie, K\"onigstuhl 17, 69117 Heidelberg, Germany
             \and
               Max-Planck-Institut f\"ur extraterrestrische Physik, Gie{\ss}enbachstra{\ss}e 1, 85748 Garching b. M\"unchen, Germany
             \and
               Instituto de Radioastronom\'ia y Astrof\'isica, Universidad Nacional Aut\'onoma de M\'exico, Antigua Carretera a P\'atzcuaro \# 8701,\\ Ex-Hda. San Jos\'e de la Huerta, Morelia, Michoac\'an, M\'exico C.P. 58089
             }

\date{Received xxx; accepted xxx}

\abstract
{} 
{We aim to understand the physical mechanisms that drive star formation in a sample of mass-complete (>10$^{9.5}M_{\odot}$) star-forming galaxies (SFGs) at 1.2 $\leq z$ < 1.6.}
{We selected SFGs from the COSMOS2020 catalog and applied a $uv$-domain stacking analysis to their archival Atacama Large Millimeter/submillimeter Array (ALMA) data. Our stacking analysis provides precise measurements of the mean molecular gas mass and size of SFGs down to a stellar mass of $M_{\star}$ $\sim$10$^{9.5}M_{\odot}$, even though at these stellar mass galaxies on the main sequence (MS) are no longer detected individually in the archival ALMA data. We also applied an image-domain stacking analysis on their \textit{HST} $i$-band and UltraVISTA $J$- and $K_{\rm s}$-band images. This allowed us to trace the distribution of their stellar component. Correcting these rest-frame optical sizes using the $R_{\rm half-stellar-light}$-to-$R_{\rm half-stellar-mass}$ conversion at rest $5000\AA$, we obtain the stellar mass size of MS galaxies and compare them to the sizes of their star-forming component obtained from our ALMA stacking analysis.}
{Across the MS (-0.2 < $\Delta$MS=log(SFR/SFR$_{\rm MS}$) < 0.2), the mean molecular gas fraction of SFGs increases by a factor of $\sim$1.4, while their mean molecular gas depletion time decreases by a factor of $\sim$1.8. The scatter of the MS could thus be caused by variations in both the star formation efficiency and molecular gas fraction of galaxies. The mean molecular gas fraction of MS galaxies decreases by a factor of $\sim$7 from $M_{\star}$$\sim$ 10$^{9.7}M_{\odot}$ to $\sim$ 10$^{11.3}M_{\odot}$, while their mean molecular gas depletion time remains roughly the same at all stellar masses. This finding could be a hint that the bending of the MS at $z\sim$1.4 is primarily driven by variations in cold gas accretion. The majority of the galaxies lying on the MS have $R_{\rm FIR}$ $\approx$ $R_{\rm stellar}$. Their central regions are subject to large dust attenuation. Starbursts (SBs, $\Delta$MS>0.7) have a mean molecular gas fraction $\sim$2.1 times larger and mean molecular gas depletion time $\sim$3.3 times shorter than MS galaxies. Additionally, they have more compact star-forming regions ($\sim$2.5~kpc for MS galaxies vs. $\sim$1.4~kpc for SBs) and systematically disturbed rest-frame optical morphologies, which is consistent with their association with major-mergers. SBs and MS galaxies follow the same relation between their molecular gas mass and star formation rate surface densities with a slope of $\sim1.1-1.2$, that is, the so-called Kennicutt-Schmidt relation.}
{}

                
\titlerunning{Molecular gas mass and extent of galaxies across and above the main sequence}
\authorrunning{T.-M. Wang et al.}         
\maketitle

\section{Introduction}
Parts of the evolutionary history of the Universe have been revealed to us in the last decades by modern telescopes. In particular, it has been found that the cosmic star formation rate density (SFRD) increases from early cosmic times, reaches a peak at cosmic noon, i.e., $z\sim2$, and smoothly decreases by a factor of $\sim10$ by $z\sim0$ \cite[e.g.,][]{2014ARA&A..52..415M, 2017A&A...602A...5N, 2018ApJ...853..172L, 2020A&A...643A...8G, 2020ApJ...899...58L}. In the star formation rate (SFR) versus stellar mass ($M_{\star}$) plane, star-forming galaxies (SFGs) can be broadly classified into two populations. The first and largest population resides on a tight correlation between SFR and stellar mass, which accounts for about 80$\%$ of the cosmic SFRD, the so-called main sequence (MS) of SFGs \citep[e.g.,][]{2007ApJ...660L..43N, 2011A&A...533A.119E, 2014A&A...561A..86M, 2015A&A...575A..74S, 2020ApJ...899...58L, 2022ApJ...936..165L, 2022MNRAS.tmp.3001P}. The MS has a scatter of $\sim$0.3~dex and a normalization that decreases by a factor of 20 from $z\sim2$ to $z\sim0$ \citep[e.g.,][]{2015A&A...575A..74S, 2020ApJ...899...58L, 2022ApJ...936..165L, 2022MNRAS.tmp.3001P}. The MS is also found to bend at high stellar mass \citep[e.g.,][]{2015A&A...575A..74S, 2019MNRAS.490.5285P, 2020ApJ...899...58L,2021A&A...647A.123D}, but this flattening becomes less prominent with increasing redshift and almost vanishes by $z\sim2$. SFGs on the MS are thought to evolve in isolation with long star-forming duty cycles, sustained by continuous cold gas accretion \citep[e.g.,][]{2019ApJ...887..235L, 2020ARA&A..58..157T, 2022A&A...660A.142W}. The second population, known as starbursts (SBs), exhibits intense star formation activities and is offset above the MS with $\Delta$MS (log$_{10}$(SFR/SFR$_{\rm MS}$)) > 0.7 \citep[e.g.,][]{2015A&A...575A..74S, 2018ApJ...867...92S}. These galaxies make up only 5$\%$ of the SFG population and contribute 10$\%$ of the SFRD out to $z\sim$4 \citep[e.g.,][]{2011ApJ...739L..40R, 2012ApJ...747L..31S, 2013A&A...552A..44L, 2015A&A...575A..74S}. Their enhanced star formation is believed to result from mergers, interactions between galaxies, or disk instabilities of galaxies \citep[e.g.,][]{2010MNRAS.404.1355D, 2012ApJ...760...11H, 2014MNRAS.444..942P, 2018MNRAS.479..758W, 2019A&A...632A..98C, 2019MNRAS.485.5631C, 2022MNRAS.516.4922R}. Despite the well-established classification of SFGs based on their SFRs and stellar masses up to $z\sim6$, the underlying mechanisms causing the evolution of the normalization of the MS (leading to the dispersion of the MS) and the triggering of SBs remain debated. Deeper insight into the mechanisms behind these scaling relations can be gained from precise measurements of their molecular gas reservoir, which serves as the fuel for star formation.

In the local universe, carbon monoxide (CO) emission from the molecular gas is widely used to measure the molecular gas content of galaxies \citep[see][for a review]{2013ARA&A..51..207B}. However, obtaining the molecular gas mass of SFGs through CO observations at high redshifts is a challenge considering the sensitivity of even the latest (sub)millimeter and radio interferometers, e.g., Atacama Large Millimeter/submillimeter Array (ALMA), Northern Extended Millimeter Array (NOEMA), and Karl G. Jansky Very Large Array (JVLA). To alleviate this limitation, a new approach to measure gas mass for high-redshift galaxies from dust continuum observations has been developed in recent years. This approach relies on standard gas-to-dust mass relations and well-calibrated dust mass measurements, which can be achieved by using either multiwavelength dust spectral energy distribution (SED) fits \citep[e.g.,][]{2012ApJ...760....6M, 2014A&A...562A..30S, 2019A&A...621A..51H, 2021ApJ...921...40K, 2021ApJ...908...15S} or a single Rayleigh--Jeans (RJ) flux density conversion \citep[e.g.,][]{2014ApJ...783...84S, 2016ApJ...820...83S, 2019ApJ...887..235L,  2020ApJ...892...66M, 2020MNRAS.494..293M, 2022MNRAS.510.3734D, 2022A&A...660A.142W}. A plethora of literature studies have since used this approach to measure the molecular gas mass of massive SFGs up to $z\sim$5 \citep[e.g.,][]{2016ApJ...820...83S, 2019ApJ...887..235L, 2020ARA&A..58..157T, 2021ApJ...908...15S,2022A&A...660A.142W}. Among these, that of \citet{2022A&A...660A.142W} is the first to statistically constrain (through stacking) the molecular gas content of a mass-complete sample of MS galaxies down to $10^{10}\,M_\odot$ and up to $z\sim3.5$, thus avoiding any selection bias that could potentially affect previous studies. With this approach, they accurately measured that the mean molecular gas fraction, i.e., $\mu_{\rm mol}=M_{\rm mol}/M_{\star}$, of MS galaxies decreases with increasing stellar mass and decreases by a factor of $\sim20$ from $z\sim3$ to $z\sim0$. Additionally, they found that the molecular gas depletion time (i.e., $\tau_{\rm mol}=M_{\rm mol}/{\rm SFR}$) of MS galaxies remains roughly constant at $z>0.5$ with a value of $300-500\,$Myr, and it only increases by a factor of $\sim3$ from $z\sim0.5$ to $z\sim0$. This suggests that to the first order, the molecular gas content of MS galaxies regulates their star formation across cosmic time, while variation of their star formation efficiency plays a secondary role. 

Although the study of \citet{2022A&A...660A.142W} has provided us with an accurate estimate of the redshift evolution of the molecular gas content of MS galaxies as a whole population, they did not study how the molecular content of SFGs evolves across and well above the MS. Therefore, while it is generally accepted that SBs partly have a higher star formation efficiency and partly a higher molecular gas content than MS galaxies \citep[e.g.,][]{2010MNRAS.407.2091G, 2019ApJ...887..235L, 2020ARA&A..58..157T}, the exact balance between these two distinct effects is still uncertain, mainly because current studies on the evolution of the molecular gas content of SFGs with $\Delta$MS are still not based on mass-complete samples and are thus potentially affected by selection biases. As a consequence, we are missing a complete picture of the physical mechanisms that drive and regulate the dispersion of $\sim0.3\,$dex of the MS and that drive galaxies well above the MS. To make progress on these topics, a detailed study of the evolution of the molecular gas content of a mass-complete sample SFGs as a function of $\Delta$MS is needed.

To gain a better understanding of the mechanisms that drive the evolution of SFGs, it is also essential to accurately measure the extent of their gas reservoir. Indeed, such measurements allow us to investigate the fundamental scaling relationship between the SFR and gas surface densities of SFGs, the so-called Kennicutt-Schmidt (KS) relation \citep[][]{1998ApJ...498..541K}. \citet{2022A&A...660A.142W} found, for example, that the moderate redshift evolution of the molecular gas depletion time of MS galaxies could be predicted from the evolution of their molecular gas surface density and a universal KS relation with a slope of $\sim1.13$. In addition, some observations suggest that the most extreme SBs follow a different KS relation, characterized by a slope of $\sim1.13$ but with a normalization $\sim10$ times higher than that of MS galaxies \citep[e.g.,][]{2010ApJ...714L.118D, 2010MNRAS.407.2091G}. This latter result is, however, still debated because it remains unclear whether the KS relation is truly bimodal (MS vs. SB) or whether it evolves smoothly with $\Delta$MS. To make further progress on this topic, a detailed study of the evolution of the molecular gas mass and size of a mass-complete sample SFGs as a function of $\Delta$MS is needed.

In this work, we thus derived the mean molecular gas mass and extent of SFGs across the MS at $M_{\star}$ > 10$^{9.5}M_{\odot}$ and $1.2\leq z \leq 1.6$ using a stacking analysis. This redshift range was chosen as it yields detection with the highest significance level in \citet{2022A&A...660A.142W}, i.e., we can further dissect the mean molecular gas mass of the SFGs in different $\Delta$MS. We selected our mass-complete sample of SFGs in the Cosmic Evolution Survey (COSMOS) field using their latest optical-to-near-infrared catalog \citep[i.e., the COSMOS2020 catalog;][]{2021AAS...23721506W} and the super-deblended far-infrared-to-millimeter catalog of \citet[][]{2018ApJ...864...56J}. These SFGs were then subdivided into different stellar mass and $\Delta$MS bins. In each bin, we performed a $uv$-domain stacking analysis on the ALMA band 6 and 7 dataset and measured their mean molecular gas mass and size. This allowed us to constrain the KS relation for a mass-complete ($>10^{9.5}\,M_\odot$) sample of galaxies probing a wide range of $\Delta$MS for the first time and thereby gain insights into the mechanisms that regulate the dispersion of the MS and that drive galaxies well above the MS. Finally, we compared the dust-obscured star-forming sizes of our MS galaxies to the size of their stellar component using data from the \textit{Hubble} Space Telescope (\textit{HST}) and UltraVISTA. This allowed us to better understand the processes leading to the structural evolution of SFGs \citep[e.g.,][]{2020ApJ...892....1W}. We verified, in particular, if the recent finding of compact star-forming regions relative to the mass-size relation of late type galaxies with ALMA \citep[e.g.,][]{2015ApJ...810..133I, 2015ApJ...799...81S, 2016ApJ...833..103H, 2017ApJ...850...83F, 2018A&A...616A.110E, 2019ApJ...879...54L, 2019ApJ...882..107R, 2019ApJ...877L..23P, 2020ApJ...888...44C, 2020A&A...635A.119C, 2020A&A...643A..30F, 2020ApJ...901...74T, 2021MNRAS.508.5217P, 2022A&A...659A.196G, 2022A&A...660A.142W} is due to dust attenuation biasing current measurements of the mass-size relation at high redshift or not.

Our paper is organized as follows. Sect.~\ref{sec:data} introduces our mass-complete sample of SFGs and their ALMA, \textit{HST}, and UltraVISTA observations; Sect.~\ref{sec:method} details the stacking method used to measure their mean molecular gas masses, star-forming sizes, and stellar sizes; Sect.~\ref{sec:results} presents our results, and Sect.~\ref{sec:discussion} discusses the implication of our findings; finally, Sect.~\ref{sec:summary} presents our conclusions.

We assumed a flat $\rm \Lambda$ cold dark matter cosmology with $H_{0}$ = 67.8 km s$^{-1}$ Mpc$^{-1}$, $\rm \Omega_{M}$ = 0.308, and $\rm \Omega_{\Lambda}$ = 0.692 \citep{2016A&A...594A..13P}. A \citet{2003PASP..115..763C} initial mass function is assumed for all stellar masses and SFRs. 

\section{Data}
\label{sec:data}
\subsection{Parent sample and ALMA data}
The latest optical-to-near-infrared catalog in the COSMOS field, i.e., the COSMOS2020 catalog \citep[][]{2021AAS...23721506W}, provides accurate photometry, photometric redshift, stellar mass, and SFR measurements for more than 966,000 sources. The astrometry of all these sources is aligned to \textit{Gaia,} and their photometry were obtained using a new photometric extraction tool (\texttt{THE FARMER}). This catalog provides highly reliable photometric redshift even for faint galaxies, that is, with a redshift precision ($\Delta z$/(1+$z$)) of 0.036 for galaxies with an $i$-band apparent magnitude between 25.0 and 27.0.

We selected galaxies at 1.2$\leq z$ <1.6 in \texttt{THE FARMER} catalog. Then, we classified galaxies into star-forming and quiescent galaxies following the standard rest-frame near-ultraviolet-$r$/$r$-$J$ selection method from \citet[][]{2013A&A...556A..55I} and selected only the SFGs. We excluded galaxies below the mass-completeness limit of the COSMOS2020 catalog, which is 10$^{8.5}M_{\odot}$ at $z\sim$1.2 and $\sim$10$^{8.7}M_{\odot}$ at $z\sim$1.6. In addition, we excluded galaxies classified as active galactic nuclei (AGNs) to avoid contamination. We excluded AGNs based on their X-ray luminosity \citep[$L_{\rm X}$ $\geq$ 10$^{42}$ erg~s$^{-1}$;][]{2004ApJS..155..271S} and the latest COSMOS X-ray catalog of \citet{2016ApJ...817...34M}. We then excluded galaxies classified as AGNs from their optical SED using the catalog of \citet[][]{2020ApJ...888...78S}. Finally, we excluded galaxies classified as radio AGNs in the VLA-COSMOS 3~GHz AGN catalog \citep[][]{2017A&A...602A...3D}. This parent sample of SFGs at 1.2 $\leq z$ < 1.6 contains 62,681 galaxies.

The A$^{3}$COSMOS dataset \citep[][]{2019ApJS..244...40L} gathers all ALMA archive projects available in the COSMOS field, making it ideal for studying the dust RJ-tail luminosity emitted by galaxies, which is a proxy of their molecular gas content. In this work, we used the A$^{3}$COSMOS data version 20220105 (Adscheid et al. in prep.), i.e., all public COSMOS ALMA datasets released by the 5 January 2022. The A$^{3}$COSMOS database contains the ALMA calibrated measurement sets, the cleaned images, and a value-added catalogue, which gathers all the galaxies individually detected in these images \citep[i.e., with a signal-to-noise ratio > 4.35 following][]{2019ApJS..244...40L}. As stressed in \citet[][]{2022A&A...660A.142W}, the definition of weights on visibilities after ALMA cycle 2 is different from the definition of the previous ALMA cycle. Therefore, we only used the ALMA data from band 6 ($\sim$243 GHz) and band 7 ($\sim$324 GHz) that were observed after ALMA cycle 2. We cross-matched our parent sample with the A$^{3}$COSMOS database, and excluded galaxies that were not covered by any ALMA observations, i.e., those located in regions with a primary beam response under 0.5. 
This reduces our sample to 1,971 galaxies.
Finally, in order to avoid contamination from bright nearby sources during our stacking analysis, we excluded galaxy pairs ($<2\farcs0$) with $S_{\rm ALMA}^1/S_{\rm ALMA}^2>2$ or $M_{\star}^1/M_{\star}^2>3$\footnote{1 and 2 denote the brighter and fainter objects in a pair system.}. There are 1,688 galaxies in this final sample.

\begin{figure}
\centering
\includegraphics[width=\columnwidth]{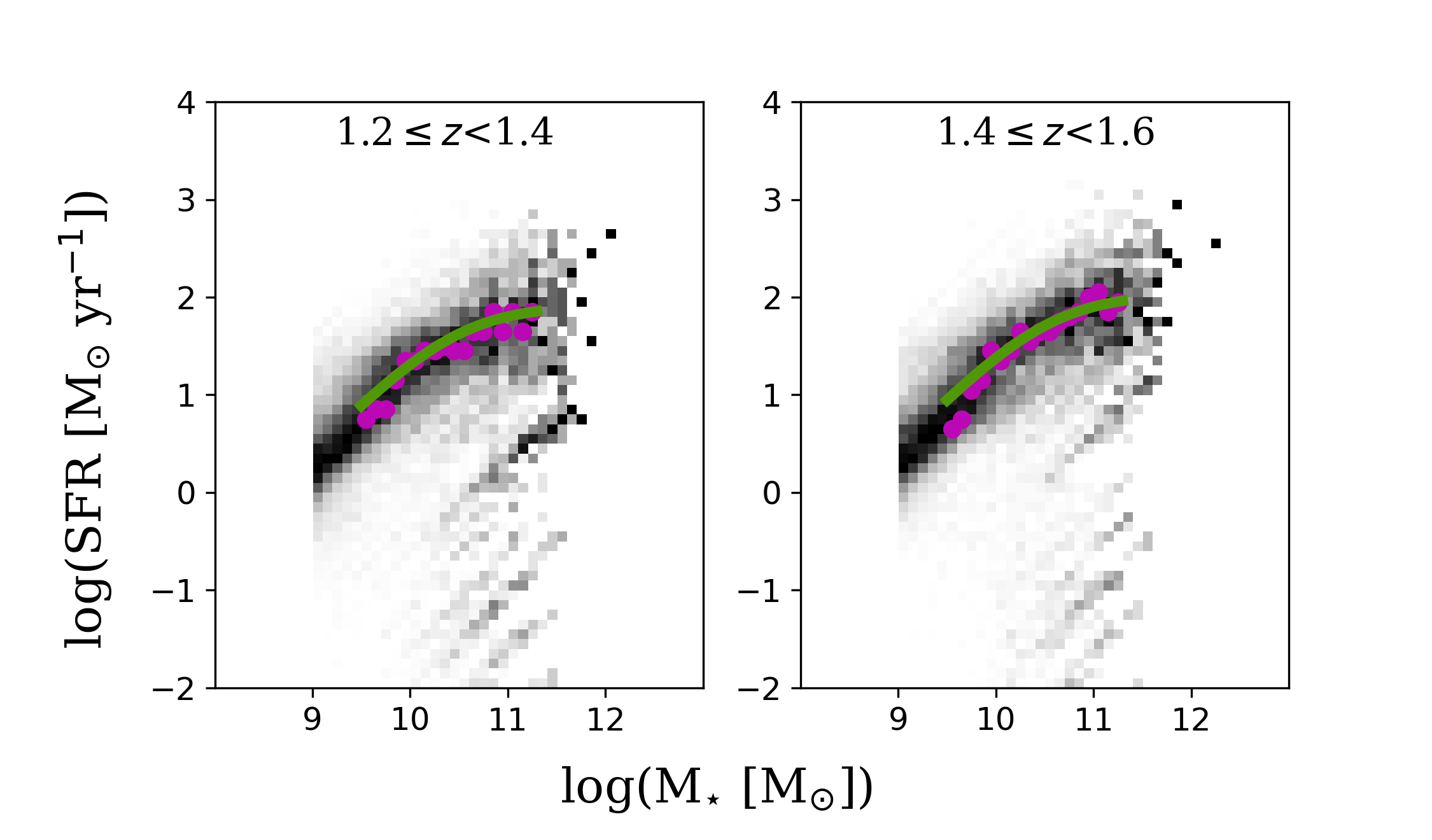}
\caption{Number density of galaxies from COSMOS2020 catalog in SFR-$M_{\star}$ plane, with SFRs determined by the approach used in our study. The darkest color indicates the highest number density of galaxies for a given stellar mass, which is further highlighted by purple circles in the stellar mass range of 10$^{9.5}M_{\odot}$$\leq M_{\star}$<10$^{11.3}M_{\odot}$. These number density peaks correspond to the MS for our SFR tracer and can be compared to the MS from \citet[][green solid lines]{2020ApJ...899...58L}, where the SFRs were measured from three~GHz radio-continuum observations.}
\label{fig:SFR_indicator}
\end{figure}
The SFR of our galaxies was obtained following \citet{1998ARA&A..36..189K} for a \citet{2003PASP..115..763C} initial mass function,
\begin{equation}
{\rm SFR}_{\rm UV+IR}[{\rm M_{\odot}\,yr^{-1}}]=1.09 \times 10^{-10} (L_{\rm IR}[{\rm L_{\odot}}]+3.3 \times L_{\rm UV}[{\rm L_{\odot}}]),
\end{equation}
where the rest-frame $L_{\rm UV}$ at 2300$\AA$ was taken from the COSMOS2020 catalog, and the rest-frame $L_{\rm IR}$ = $L({\rm 8-1000 \,\mu m})$ was calculated by following the ladder of SFR indicator as advocated by \citet[][]{2011ApJ...738..106W}; i.e., we used the best infrared luminosity indicator available for each galaxy. To this end, we used the mid-infrared (MIR) and far-infrared (FIR) photometry from the super-deblended catalog \citep[][]{2018ApJ...864...56J}, which provides \textit{Spitzer} and \textit{Herschel} photometries without contamination from nearby sources by iteratively deblending source emission with prior information, SED prediction, and residual source detection techniques. For galaxies with FIR and MIR photometry, their $L_{\rm IR}$ was obtained by fitting their MIR and/or FIR SED with the template of \citet{2001ApJ...556..562C} (see details in Sect.~2.2 of \cite{2022A&A...660A.142W}). For galaxies without any MIR or FIR photometry, their SFRs were taken directly from the COSMOS2020 catalog. We note that \cite{2022A&A...660A.142W} have verified that the SFRs obtained from UV+IR measurements agree with the COSMOS2020 SFRs at intermediate SFRs (see their Fig.~2.), i.e., where the MIR/FIR detection rate of galaxies starts to decrease. From the SFR and stellar mass of each of our galaxies, we measured their offset from the MS, i.e., $\Delta$MS = ${\rm log(SFR}(z,M_{\star})/{\rm SFR}_{\rm MS}(z,M_{\star}))$, using the MS calibration of \citet{2020ApJ...899...58L}. 

The SFR indicator used in our study differs from that of \citet{2020ApJ...899...58L}, who used 3~GHz radio-continuum to measure the SFR of their SFG. In Fig.~\ref{fig:SFR_indicator}, we show a test we did to determine whether these different SFR indicators could introduce bias into our study, and in particular lead to different definitions of the MS. To this end, we selected all $M_{\star}\geq$10$^{9.0}M_{\odot}$ galaxies with $1.2\leq z<1.6$ in the COSMOS2020 catalog and calculated their SFR based on the ladder of SFR indicators used in our study. At 10$^{9.5}M_{\odot}$$\leq M_{\star}$<10$^{11.3}M_{\odot}$, we then defined the peak of the number density of galaxies at each stellar mass (purple circles in Fig.~\ref{fig:SFR_indicator}) as "our" MS. Finally, we compared our MS with that of \citet[][green solid lines in Fig.~\ref{fig:SFR_indicator}]{2020ApJ...899...58L} and found an average offset of 0.02~dex and no offset greater than 0.2~dex in any of our specific stellar mass bins. This demonstrates that the use of the MS from Leslie et al. is appropriate and leads to accurate measurements of $\Delta$MS.

\subsection{\textit{HST} and UltraVISTA data}
We studied the size of the stellar component of our galaxies using their optical-to-near-infrared emission. First, we used the \textit{HST}/Advanced Camera for Surveys (ACS) $i$-band images of \citet[][]{2007ApJS..172..196K} and \citet[][]{2010MNRAS.401..371M}. The \textit{HST}/ACS $i$ band has a wavelength centered at 8,333~$\AA$ and a PSF full width at half maximum ($FWHM$) of 0.095 arcsec. At the redshift of our galaxies, this corresponds to the rest-frame wavelength of $\sim$3,470~$\AA$ and thus traces the near UV emitted by young O/B stars. Second, we used the Ultra Visible and Infrared Survey Telescope for Astronomy (UltraVISTA) $J$- and $K_{\rm s}$-band images of \citet[][]{2012A&A...544A.156M}. The UltraVISTA $J$ and $K_{\rm s}$ bands have wavelengths centered on 12,525 and 21,557 $\AA$. At the redshift of our galaxies, these images correspond to the rest-frame wavelengths of $\sim$5,210~$\AA$ and $\sim$8,980~$\AA$ and thus trace the bulk of the stellar population of these galaxies. Because VISTA is a ground-based telescope, there is variation of the seeing across the mosaics. However, \citet[][]{2012A&A...544A.156M} stressed that such seeing variation is small ($\sim$0.05 arcsec) compared to the average seeing. In our paper, we thus used the PSF $FWHM$s of \citet[][]{2012A&A...544A.156M}, which are 0.77 and 0.78 arcsec for the $J$ and $K_{\rm s}$ bands, respectively. 

\section{Method}
\label{sec:method}
The high-sensitivity of ALMA has revolutionized our ability to measure the gas content (from RJ dust continuum emission) of high-stellar-mass, high-$\Delta$MS, and low-redshift galaxies. However, with the time restriction of each ALMA project, galaxies at high redshift, low stellar mass, and low $\Delta$MS remain difficult to detect individually with ALMA (e.g., see Table 1 of \citet[][]{2022A&A...660A.142W}). To statistically measure the mean value of our mass-complete sample of SFGs at $z\sim$1.4, we therefore used a stacking analysis. However, stacking the A$^3$COSMOS dataset is a challenge because the different ALMA projects were observed at different frequencies and spatial resolutions \citep[e.g.,][]{2022A&A...660A.142W}. On the contrary, stacking the \textit{HST} or UltraVISTA datasets is straightforward because each of these datasets corresponds to only one frequency and one angular resolution.
Below, we describe the steps to perform the $uv$-domain stacking analysis on our ALMA data and the steps to perform the image-domain stacking analysis on the \textit{HST} and UltraVISTA data.

\subsection{The stacking analysis}
We followed the same approach as \citet[][]{2022A&A...660A.142W} to perform the $uv$-domain stacking analysis of our ALMA dataset. Details regarding these steps can be found in their Sects.~3.1 and 3.2. We first scaled the observed ALMA visibility amplitudes of each galaxy to its rest-frame luminosity at 850\,$\mu$m (i.e., $L_{\rm 850}^{\rm rest}$) using the SED templates of \citet{2012ApJ...757L..23B} and the Common Astronomy Software Applications package \citep[CASA;][]{CASA} tasks \texttt{gencal} and \texttt{applycal}. In practice, this scaling was performed using both the MS and SB galaxy SED templates of \citet{2012ApJ...757L..23B}. Then, for a galaxy with $\Delta$MS <0.4, we kept the scaling obtained with the MS galaxy SED template; for a galaxy with $\Delta$MS > 0.7, we kept that measured with the SB galaxy SED template; and, finally, for a galaxy with 0.4 < $\Delta$MS < 0.7, we used a linear interpolation in the log-space between the MS and SB galaxy SED template scalings. For each galaxy, we then shifted the phase center of its visibilities to its coordinates by using a CASA based package: the \texttt{STACKER} \citep{2015MNRAS.446.3502L}. The visibilities of the galaxies in each $M_{\star}$--$\Delta$MS bin were then stacked together using the CASA task \texttt{concat}. The cleaned image was generated from the stacked measurement set with the CASA task \texttt{tclean} using Briggs $n=2$ weighting and cleaning the image down to $3\sigma$. Finally, we measured the stacked $L_{\rm 850}^{\rm rest}$ and its beam-deconvolved distribution with the CASA task \texttt{uvmodelfit} ($uv$-domain) and the Python Blob Detector and Source Finder (\texttt{PyBDSF}) package \citep{2015ascl.soft02007M} (image-domain). Both \texttt{uvmodelfit} and \texttt{PyBDSF} fit a single Gaussian component to the source. We express the mean size of the stacked galaxies in the form of circularized radii, $R^{\rm circ}_{\rm eff}$:
\begin{equation}
R_{\rm eff}^{\rm circ} = R_{\rm eff} \times \sqrt{\frac{b}{a}},
\end{equation}
where $b/a$ is the axis ratio measured with \texttt{uvmodelfit} or \texttt{PyBDSF} and $R_{\rm eff}$ is the effective radius (following \citet{2017ApJ...839...35M} and \citet{2019A&A...625A.114J}, $R_{\rm eff}\approx FWHM_{\rm Gaussian}/2.43$, where $FWHM_{\rm Gaussian}$ is the beam deconvolved $FWHM$ of the single Gaussian component fitted to the source).

We followed the standard resampling method from \citet[][]{2022A&A...660A.142W} to derive the uncertainties in these $L_{\rm 850}^{\rm rest}$ and size measurements (see their Sect.~3.2). This gives us the uncertainties from both the instrumental noise and the intrinsic distribution of the stacked population. These uncertainties are marked as thin error bars in the following figures. We also measured the uncertainties from the instrumental noise alone by simulating galaxies with the same $L_{\rm 850}^{\rm rest}$ and size measured in each stacked $M_{\star}$--$\Delta$MS bin. These simulated galaxies were placed at $3\farcs0$ from the phase center in each realization of our resampling method. Then, by measuring the physical properties of these simulated galaxies, we obtained the uncertainties from the instrumental noise alone. This uncertainty is marked by thick error bars in the following figures. 

To obtain the mean molecular gas mass of each $M_{\star}$--$\Delta$MS bin, we first converted their stacked $L_{\rm 850}^{\rm rest}$ to dust mass. This was performed by scaling the MS and SB galaxy SED templates of \citet{2012ApJ...757L..23B} to the stacked $L_{\rm 850}^{\rm rest}$, calculating the corresponding $L_{\rm IR}$, and applying the $L_{\rm IR}$-to-dust mass conversion provided by \citet{2012ApJ...757L..23B} for each of these SED templates. Then, for bins with $\Delta$MS <0.4, we kept the dust mass measured with the MS galaxy SED template; for bins with $\Delta$MS > 0.7, we kept that measured with the SB galaxy SED template; and, finally, for bins with 0.4 < $\Delta$MS < 0.7, we used a linear interpolation in the log-space between the MS and SB galaxy SED template measurements. Because these dust mass measurements considered the $\Delta$MS of each bin, it takes into account the different dust grain temperatures within MS and SB galaxies. We then converted these dust masses into molecular gas masses using the gas-to-dust mass ratio versus metallicity relation of \cite{2011ApJ...737...12L}, i.e., $\delta_{\rm GDR}$ method:\begin{equation}
{\rm log}(M_{\rm mol})={\rm log}(M_{\rm dust}) + 9.4 - 0.85\times(\rm 12+log(O/H)),
\end{equation}
where the gas-phase metallicity, 12+log(O/H), was obtained following the redshift- and stellar mass-dependent relation given in \citet{2019ApJ...887..235L}:
\begin{equation}
{\rm 12+log(O/H)}=
\begin{cases}
a\hspace{3.0cm}  {\rm if\,log}(M_{\star}/M_{\odot})\geq b(z),\\
a-0.087\times ({\rm log}(M_{\star}/M_{\odot})-b(z))^{2},\hspace{0.7cm} {\rm else},
\end{cases}
\end{equation}
where $a$=8.74 and $b(z)$=10.4+4.46$\times$log(1+$z$)-1.78$\times$(log(1+$z$))$^{2}$.
\\

As mentioned above, images in the COSMOS field obtained by the \textit{HST} or UltraVISTA telescope have uniform sensitivity and PSF. In these cases, stacking in the image domain is straightforward and appropriate. For each telescope, we stacked the images using the noise-weighted method:
\begin{equation}
I_{{\rm stack}}=\frac{\Sigma\,I_{i}\,\sigma^{-2}_{i}}{\Sigma\sigma^{-2}_{i}},
\end{equation}
where $I_{{\rm stack}}$ is the pixel intensity in the stacked image, $I_{i}$ is the pixel intensity of the $i$-th galaxies, and $\sigma_{i}$ is the pixel noise of the $i$-th galaxies. We compare the radial profile of the stacked ALMA, \textit{HST}, and UltraVISTA images in Sect.~\ref{sect:size_morphology}. 

\begin{figure}
\centering
\includegraphics[width=\columnwidth]{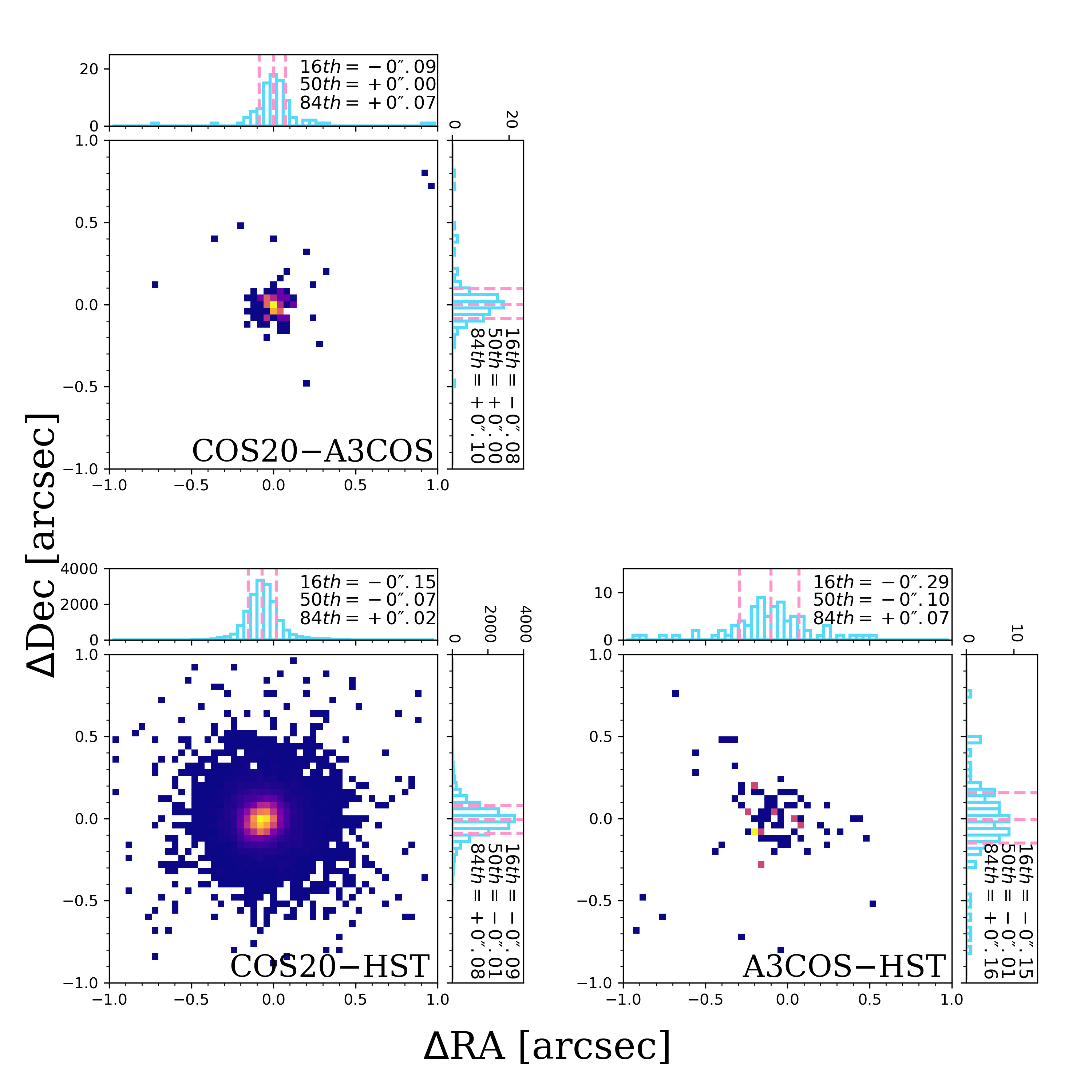}
\caption{Astrometric offsets between COSMOS-2020 \citep[][]{2021AAS...23721506W}, \textit{HST}-ACS-COSMOS \citep[][]{2007ApJS..172..219L}, and A$^{3}$COSMOS catalogs \citep[][]{2019ApJS..244...40L}. Each panel shows the R.A. and Dec. offset distribution for sources presented in both catalogs. The top and right axes of each panel display the histograms of these R.A. and Dec. offsets. Pink dashed vertical lines are the 16th, 50th, and 84th percentiles of each distribution.}
\label{fig:astrometry}
\end{figure}
\begin{figure}
\centering
\includegraphics[width=\columnwidth]{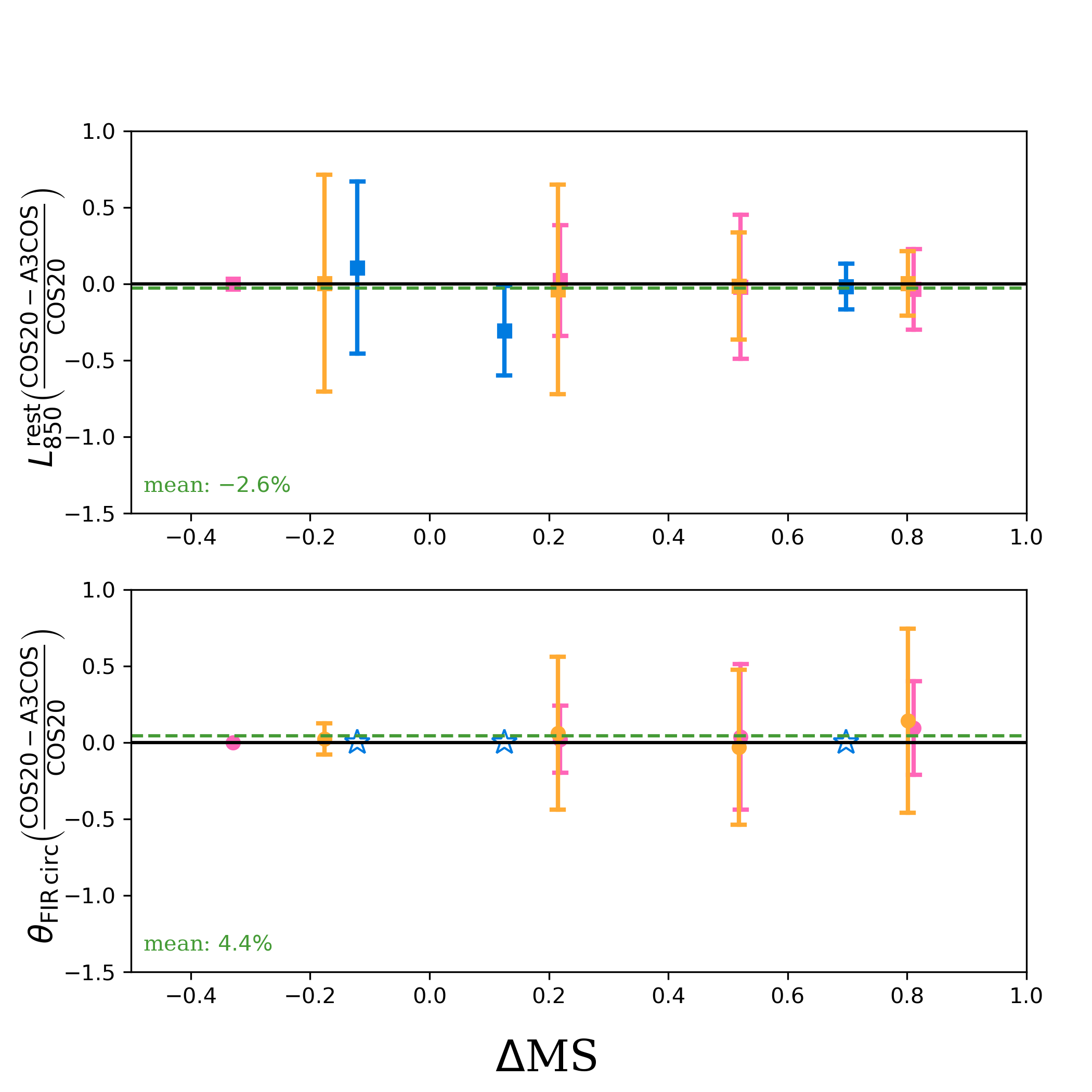}
\caption{Relative differences on rest-frame 850$\mu$m luminosity ($L^{\rm rest}_{850}$) and FIR beam-deconvolved size of the stacked galaxies based on astrometry from the COSMOS-2020 catalog \citep[][]{2021AAS...23721506W} or A$^{3}$COSMOS catalog \citep[][]{2019ApJS..244...40L}. Squares show the relative difference on the stacked $L^{\rm rest}_{850}$. Circles display the relative difference on the FIR size for the resolved stacked galaxies, while stars show the bins where the stacks are spatially unresolved (i.e., point sources). Symbols are color-coded by stellar mass, i.e., pink for 10$^{11}\leq M_{\star}/{\rm M}_{\odot}<10^{12}$, orange for 10$^{10.5}\leq M_{\star}/{\rm M}_{\odot}<10^{11}$, and blue for 10$^{10}\leq M_{\star}/{\rm M}_{\odot}<10^{10.5}$. The green dashed lines are the average uncertainties in each panel, while values are given on the bottom left of each panel. Overall, using either optically-based or (sub)millimeter-based positions in stacking analysis yields consistent results.}
\label{fig:diff_coord}
\end{figure}

\subsection{Uncertainties from the mismatch of position and selection of astrometry}
\label{sect:uncertainties}
The use of accurate galaxy coordinates is critical for obtaining high-quality results from our stacking analysis.
In Fig.~\ref{fig:astrometry}, we compare the astrometry offset between sources in the COSMOS2020 and \textit{HST}-ACS-COSMOS catalog and show a 0.07 arcsec offset in R.A. This systematic offset is not due to an intrinsic mismatch between the \textit{HST} and COSMOS2020 emitting regions of our galaxies, but rather a misalignment between the astrometric references of the two catalogs. Despite being small, this misalignment is still larger than the PSF size of the stacked \textit{HST} image and cannot be neglected. To account for this, a 0.07 arcsec offset in R.A. was taken into consideration when we performed the image-domain stacking analysis on the \textit{HST} $i$-band image.

A possible intrinsic mismatch of $\sim 0\farcs2$ between the optical and FIR emitting regions of SFGs was reported in some literature studies \citep[e.g.,][]{2018A&A...616A.110E}. This would translate into a dispersion between the FIR and optical positions of SFGs and could lead to artificially large FIR sizes when stacking the FIR data using optical positions. We investigated the existence of such systematic positional mismatch by comparing the astrometry of the COSMOS2020 catalog (optical-based; positions used in our ALMA stacking analysis) and A$^{3}$COSMOS catalog \citep[][(sub)millimeter-based]{2019ApJS..244...40L} in Fig.~\ref{fig:astrometry}. We find that the positional dispersions between these catalogs is relatively small compared to the beam size of our stacked ALMA dataset (i.e., dispersion < 1/3 $FWHM$ of the typical ALMA stacked beam). We then studied the impact of this dispersion on our results further by repeating our stacking analysis but limiting it to the source individually detected by ALMA and using their A$^{3}$COSMOS position instead
of their COSMOS2020 position. We find that using the A$^{3}$COSMOS positions changes by their measured stacked luminosities and sizes (Fig.~\ref{fig:diff_coord}) at most 5$\%,$  which is relatively little compared to our uncertainties. Based on this finding, we conclude that the mismatch between the optical-based position and (sub)millimeter-based position would only have a minor effect on the FIR luminosities and sizes inferred from our stacking analysis.

\begin{figure*}
\centering
\includegraphics[width=1.0\textwidth]{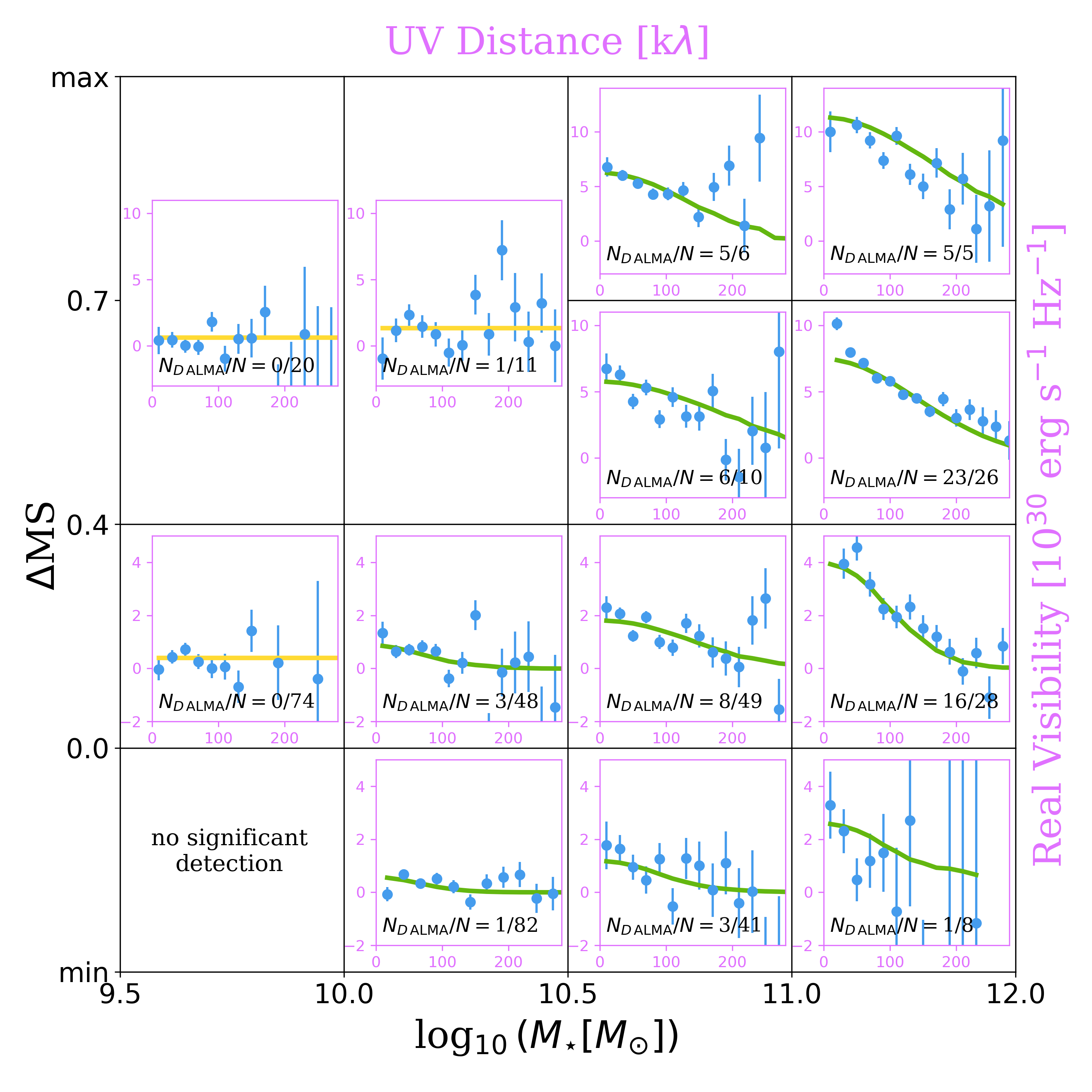}
\caption{Results in $uv$-domain of our ALMA stacking analysis. In each panel, a single component model (green solid line for Gaussian profile and yellow solid line for point source) has been fit to the mean visibility amplitudes (blue filled circles) using the CASA task \texttt{uvmodelfit}. The maximun values of $\Delta$MS from the lowest to highest stellar mass bins are 1.3, 0.9, 1.0, and 1.1; and the minimum values from the lowest to highest stellar mass bins are -0.7, -0.7, -0.7, and -0.6, respectively. The number of individually detected galaxies in the ALMA archival images ($N_{D\,{\rm ALMA}}$) and the number of stacked galaxies ($N$) is also reported in each panel.}
\label{fig:uvmodel}
\end{figure*}
\begin{sidewaystable*}
\footnotesize
\centering
\caption{Molecular gas mass and size properties of the stacked galaxies.}
\label{tab:sm}
\begin{tabular}{cccccccccccccccc}
\hline
$M_\star$ & $\Delta$MS & $N$ & $N_{D\,{\rm ALMA}}$ & $\langle z \rangle$ & $\langle M_\star\rangle$ & $\langle$$\Delta$MS$\rangle$ & $\langle$$L^{\rm rest}_{850-uv}$$\rangle$ & S/N$_{\rm peak}$ & $M_{\rm mol-uv}$ & $M_{\rm mol-py}$ & $\theta_{\rm beam}^{\rm circ}$ & $\theta_{\rm decon.-uv}^{\rm circ}$ & $\theta_{\rm decon.-py}^{\rm circ}$ & $R_{\rm \rm eff-uv}^{\rm circ}$ & $R_{\rm \rm eff-py}^{\rm circ}$ \\
 & & & & & log$_{10}$\,M$_{\odot}$ & & 10$^{30}$\,erg\,s$^{-1}$ & & log$_{10}$\,M$_{\odot}$ & log$_{10}$\,M$_{\odot}$ & arcsec & arcsec & arcsec & kpc & kpc\\
 &&&&&&&Hz$^{-1}$&&&&&\\
 (1) & (2) & (3) & (4) & (5) & (6) & (7) &  (8) & (9) & (10) & (11) & (12) & (13) & (14) & (15) & (16) \\
\hline
11.0$\leq$ log$_{10}\ M_{\star}$ <12.0 & 0.7$\leq$ $\Delta$MS <1.1 & 5 & 5  & 1.38 & 11.25 & 0.81 & 11.4$_{\pm 2.6}$ & 27 & 11.0$_{\pm 0.1}$ & 11.0$_{\pm 0.1}$ & 0.61 & 0.26 & 0.32 & 0.9$_{\pm 0.3}$ & 1.1$_{\pm 0.3}$\\
 & 0.4$\leq$ $\Delta$MS <0.7 & 26 & 23 & 1.40 & 11.29 & 0.54 & 7.5$_{\pm 1.0}$ & 46 & 10.8$_{\pm 0.1}$ & 10.8$_{\pm 0.1}$ & 0.51 & 0.31  & 0.37 & 1.1$_{\pm 0.1}$ & 1.3$_{\pm 0.1}$\\
 & 0.0$\leq$ $\Delta$MS <0.4 & 28 & 16 & 1.36 & 11.27 & 0.20 & 4.0$_{\pm 0.9}$ & 24 & 10.7$_{\pm 0.1}$ & 10.8$_{\pm 0.1}$ & 0.52 & 0.48  & 0.51  & 1.7$_{\pm 0.3}$ & 1.8$_{\pm 0.3}$\\
 & -0.6$\leq$ $\Delta$MS <0.0 & 8 & 1 & 1.43 & 11.26 & -0.30 & 2.6$_{\pm 1.9}$ & 17 & 10.6$_{\pm 0.3}$ & 10.6$_{\pm 0.3}$ & 0.94 & 0.44 & 0.49 & 1.6$_{\pm 1.2}$ & 1.7$_{\pm 1.2}$\\
\hline
10.5$\leq$ log$_{10}\ M_{\star}$ <11.0 & 0.7$\leq$ $\Delta$MS <1.0 & 6 & 5  & 1.45 & 10.83 & 0.80 & 6.3$_{\pm 1.6}$ & 21 & 10.8$_{\pm 0.1}$ & 10.8$_{\pm 0.1}$ & 0.68 & 0.20 & 0.34 & 0.7$_{\pm 0.6}$ & 1.2$_{\pm 0.6}$\\
 & 0.4$\leq$ $\Delta$MS <0.7 & 10 & 6 & 1.31 & 10.78 & 0.50 & 5.8$_{\pm 1.2}$ & 17 & 10.7$_{\pm 0.1}$ & 10.8$_{\pm 0.1}$ & 0.66 & 0.27 & 0.35 & 0.9$_{\pm 0.4}$ & 1.2$_{\pm 0.4}$\\
 & 0.0$\leq$ $\Delta$MS <0.4 & 49 & 8 & 1.35 & 10.74 & 0.17 & 1.8$_{\pm 1.2}$ & 16 & 10.5$_{\pm 0.3}$ & 10.5$_{\pm 0.3}$ & 0.69 & 0.40 & 0.48  & 1.4$_{\pm 0.3}$ & 1.7$_{\pm 0.3}$\\
 & -0.7$\leq$ $\Delta$MS <0.0 & 41 & 3 & 1.34 & 10.71 & -0.22 & 1.2$_{\pm 0.8}$ & 5 & 10.3$_{\pm 0.3}$ & 10.5$_{\pm 0.2}$ & 0.70 & 0.58 & 0.86  & 2.0$_{\pm 1.9}$ & 3.0$_{\pm 2.0}$\\
\hline
10.0$\leq$ log$_{10}\ M_{\star}$ <10.5 & 0.4$\leq$ $\Delta$MS <0.9 & 11 & 1  & 1.39 & 10.19 & 0.72 & 1.4$_{\pm 0.4}$ & 4 & 10.3$_{\pm 0.1}$ & 10.4$_{\pm 0.1}$ & 0.56 & - & - & - & < 2.0$^{\dagger}$\\
 & 0.0$\leq$ $\Delta$MS <0.4 & 48 & 3 & 1.37 & 10.25 & 0.14 & 0.9$_{\pm 0.2}$ & 6 & 10.2$_{\pm 0.1}$ & 10.4$_{\pm 0.1}$ & 0.87 & 0.56 & 0.85  & 2.0$_{\pm 1.4}$ & 3.0$_{\pm 1.4}$\\
 & -0.7$\leq$ $\Delta$MS <0.0 & 82 & 1 & 1.38 & 10.22 & -0.28 & 0.6$_{\pm 0.3}$ & 4 & 10.1$_{\pm 0.2}$ & 10.2$_{\pm 0.2}$ & 0.73 & 0.91 & 1.00 & 3.2$_{\pm 1.6}$ & 3.5$_{\pm 1.6}$\\
\hline
9.5$\leq$ log$_{10}\ M_{\star}$ <10.0 & 0.4$\leq$ $\Delta$MS <1.3 & 20 & 0  & 1.40 & 9.69 & 0.59 & 0.6$_{\pm 0.4}$ & 4 & 10.1$_{\pm 0.3}$ & 10.2$_{\pm 0.3}$ & 1.14 & - & - & - & < 4.0$^{\dagger}$\\
 & 0.0$\leq$ $\Delta$MS <0.4 & 74 & 0 & 1.34 & 9.73 & 0.18 & 0.4$_{\pm 0.2}$ & 4 & 10.0$_{\pm 0.2}$ & 9.9$_{\pm 0.2}$ & 1.07 & -  & - & - & < 3.8$^{\dagger}$\\
 & -0.7$\leq$ $\Delta$MS <0.0 & 129 & 0 & 1.38 & 9.72 & -0.29 & < 0.3$^{\dagger\dagger}$ & - & - & < 10.0$^{\dagger\dagger}$ & 0.61 & -  & - & - & -\\
\hline
\end{tabular}
\tablefoot{(1) Stellar mass bin, (2) $\Delta$MS bin, (3) number of stacked galaxies, (4) number of individually detected stacked galaxies, (5) mean redshift, (6) mean stellar mass, (7) mean $\Delta$MS, (8) mean $L^{\rm rest}_{850}$ inferred from the $uv$-domain, (9) peak S/N on the image, (10) mean molecular gas mass inferred from the $uv$-domain, and (11) from the image-domain, (12) circularized synthesized beam $FWHM$, (13) circularized deconvolved $FWHM$ from the $uv$-domain, and (14) from the image-domain, (15) circularized half-light radii from the $uv$-domain, and (16) from the image-domain. \\$^{\dagger}$Size upper limit inferred from its circularized synthesized beam $FWHM$. \\$^{\dagger\dagger}$ $L_{850}$ and molecular gas mass upper limit inferred from the 5$\sigma$ noise on the image.}
\end{sidewaystable*}
\begin{figure*}
\centering
\includegraphics[width=1.0\textwidth]{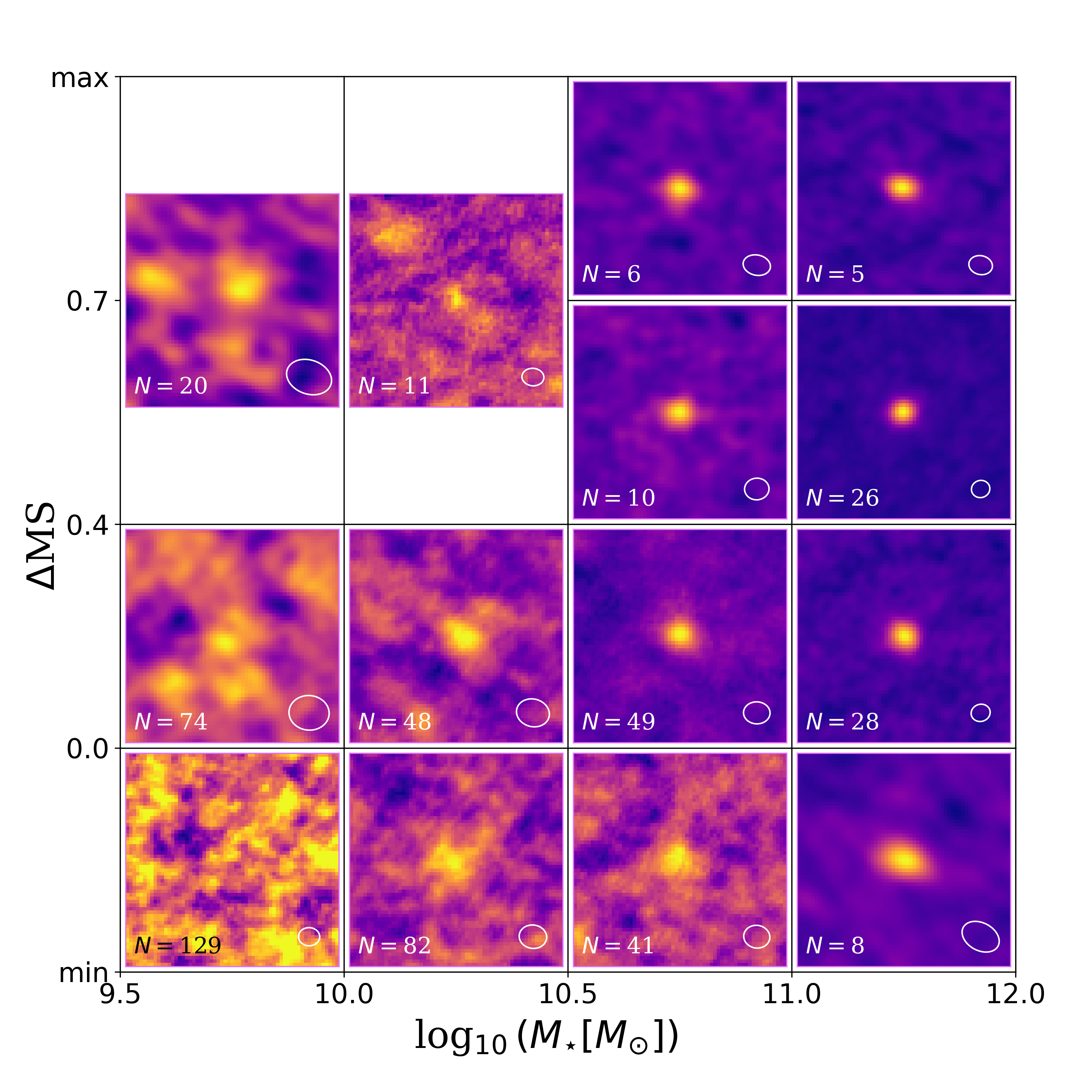}
\caption{Results in image-domain of our ALMA stacking analysis in $uv$-domain. The maximun values of $\Delta$MS from the lowest to highest stellar mass bins are 1.3, 0.9, 1.0, and 1.1; and the minimum values from the lowest to highest stellar mass bins are -0.7, -0.7, -0.7, and -0.6, respectively. Each panel has a size of 6 arcsec $\times$ 6 arcsec. The synthesized beam of the image is shown in the right-bottom corner of each panel.}
\label{fig:ALMA_img}
\end{figure*}

\section{Results}
\label{sec:results}
\subsection{Molecular gas content and star-forming size}
The results of our ALMA stacking analysis are shown in Fig.~\ref{fig:uvmodel} and Fig.~\ref{fig:ALMA_img} and are summarized in Tab.~\ref{tab:sm}. In high $M_{\star}$--$\Delta$MS bins, i.e., $M_{\star}$ > 10$^{10.5}M_{\odot}$ and $\Delta$MS > 0.4, more than 60$\%$ of the stacked sources are detected individually by ALMA. The stacks for these bins thus yield a detection with a high signal-to-noise ratio, i.e., S/N>17. The detection rate ($N_{D\,{\rm ALMA}}/N$) decreases with decreasing stellar mass and $\Delta$MS, reaching a minimum of 0$\%$ in our lowest stellar mass bins. Our stacking analysis allows, however, for the detection at reasonable significance (i.e., S/N > 4) in most of these low $M_{\star}$--$\Delta$MS bins. Yet, there is no significant detection in the lowest stellar mass and $\Delta$MS bin, i.e., $10^{9.5}M_{\odot}\leq$ $M_{\star}$ < $10^{10.0}M_{\odot}$ and $\Delta$MS < 0. For this bin, we instead infer an 5$\sigma$ upper limit on the mean molecular gas mass. 

In the $uv$ domain, most of the stacked visibilities are well fit by an elliptical Gaussian component, except for our lowest stellar mass bins and the highest $\Delta$MS bin at 10$^{10.0}M_{\odot} \leq$ $M_{\star}$ < 10$^{10.5}M_{\odot}$. Indeed, for the latter, a fit with an elliptical Gaussian component with \texttt{uvmodelfit} led to an axis ratio with a value close to 0. These bins have instead the properties of a point source, as also suggested by the relatively constant value of their visibility amplitude as a function of the $uv$ distance. For these bins, we simply fit a point-source model and conservatively used their circularized synthesized beam $FWHM$ as an upper limit to their FIR size. 

Finally, we note that the mean molecular gas mass and size measured by \texttt{uvmodelfit} in the $uv$ domain and by \texttt{PyBDSF} in the image domain are in agreement within the uncertainties (see Tab.~\ref{tab:sm}). Such an agreement is perfectly in line with the findings and simulations presented in \citet[][]{2022A&A...660A.142W}.

\begin{figure}
\centering
\includegraphics[width=\columnwidth]{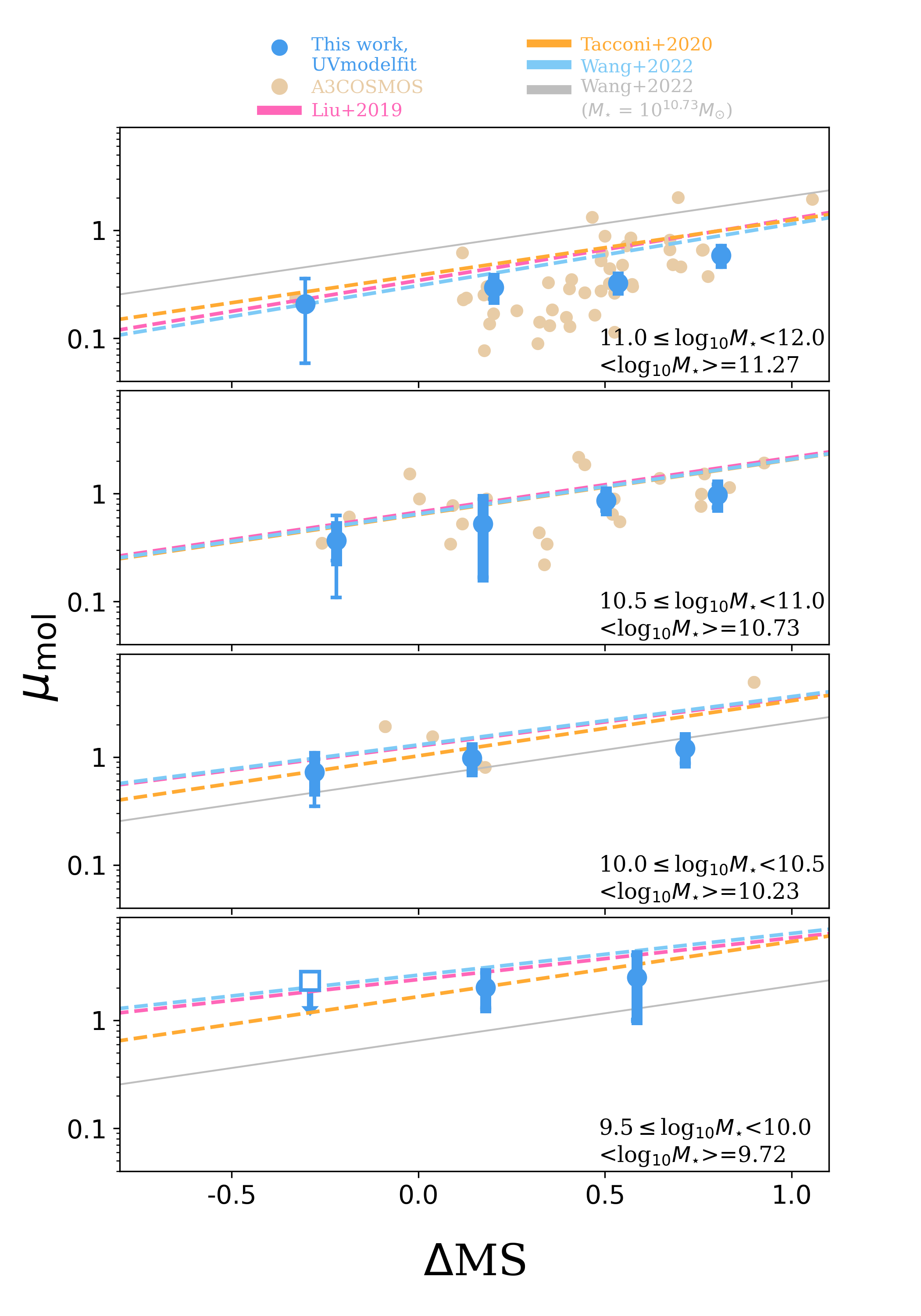}
\caption{Molecular gas fraction (i.e., $\mu_{\rm mol}$ = $M_{\rm gas}/M_{\star}$) of SFGs at $z\sim$1.4 as a function of their distance to the MS (i.e., $\Delta$MS) and their stellar mass. Blue circles show the mean molecular gas fraction from our work. The square displays the upper limit for the bin with no detection in both the $uv$ domain and image domain. The thick error bars display the instrumental uncertainty in each bin, while the thin error bars show the total uncertainty, i.e., including both the instrumental noise and the uncertainties due to the intrinsic dispersion of the molecular gas fraction within the stacked population. Salmon-colored circles represent individually detected galaxies taken from the latest A$^{3}$COSMOS catalog ($S/N>4.35$; Adscheid et al. in prep.), applying the same method used here to convert these flux densities into molecular gas mass. Dashed lines display the analytical relations of the molecular gas fraction at different stellar mass and $\Delta$MS from \citet[][pink]{2019ApJ...887..235L}, \citet[][orange]{2020ARA&A..58..157T}, and \citet[][blue]{2022A&A...660A.142W}. Gray lines show the analytical relation from \citet[][]{2022A&A...660A.142W} at $M_{\star}$ = 10$^{10.73}M_{\odot}$ as a reference, which is the same stellar mass as in the second panel.}
\label{fig:gas_frac}
\end{figure}
\begin{figure}
\centering
\includegraphics[width=\columnwidth]{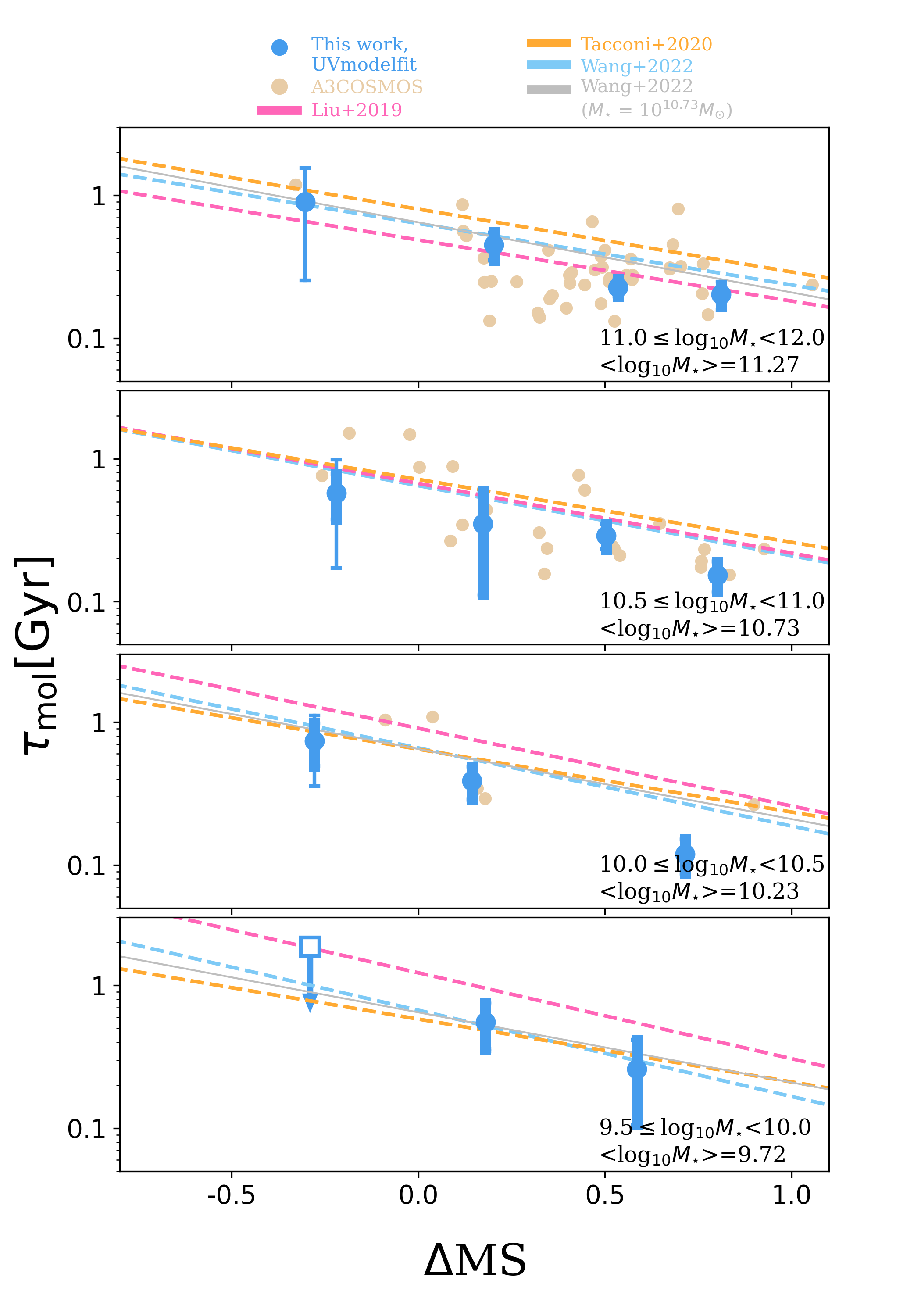}
\caption{Same as Fig.~\ref{fig:gas_frac}, but for the molecular gas depletion time (i.e., $\tau_{\rm mol}$ = $M_{\rm gas}$/SFR).}
\label{fig:depltion}
\end{figure}

\subsubsection{Molecular gas fraction and depletion time of SFGs}
\label{sect:mass_depletion}
The mean molecular gas fraction ($\mu_{\rm mol}$=$M_{\rm gas}/M_{\star}$) and mean molecular gas depletion time ($\tau_{\rm mol}$=$M_{\rm gas}/$SFR) of our stacked galaxies for different $\Delta$MSs are shown in Fig.~\ref{fig:gas_frac} and Fig.~\ref{fig:depltion}, respectively, along with the analytical relations of \citet[][]{2019ApJ...887..235L}, \citet[][]{2020ARA&A..58..157T}, and \citet[][]{2022A&A...660A.142W}.
Our measurements show that the mean molecular gas fraction of galaxies increases by a factor of $\sim$2.1 from MS galaxies ($\Delta$MS$\sim$0) to SB galaxies ($\Delta$MS$\sim$0.8), and the mean molecular gas depletion time decreases by a factor of $\sim$3.3 from MS galaxies to SB galaxies. These findings are consistent with those of \citet{2023ApJ...943...82S}, who find that the mean molecular gas fraction increases by a factor of $\sim$1.7 from MS galaxies to SB galaxies, while the mean molecular gas depletion time decreases by a factor of $\sim$3.7. These findings suggest that the variation in star formation efficiency (1/$\tau_{\rm mol}$) and the variation in gas content are roughly similar when galaxies move from being MS to SB. Meanwhile, the mean molecular gas fraction of galaxies increases by a factor of $\sim$1.4 across the MS (-0.2 < $\Delta$MS < 0.2), and the mean molecular gas depletion time decreases by a factor of $\sim$1.8 across the MS. Our findings indicate that the change of gas content and star formation efficiency play again roughly a similar role when SFGs oscillate within the MS. This disagrees with the findings of \citet{2014ApJ...793...19S}, who found that the oscillation is mainly due to variations in their gas content. This discrepancy likely comes from the fact that the "2-star formation mode" model of \citet{2014ApJ...793...19S} was fit to a relatively small number of data points, mixing different redshifts and stellar masses and restricting the variation of the depletion time of MS galaxies to their SFR. Finally, we find that the mean molecular gas fraction of galaxies decreases by a factor of $\sim$7 from $M_{\star}$ $\sim$ 10$^{9.7}M_{\odot}$ to $\sim$ 10$^{11.3}M_{\odot}$, while their mean molecular gas depletion time does not change with stellar mass. Our results support the scenario that the bending of the MS at $z\sim$1.4, which at this redshift appears at $M_{\star}$ > 10$^{10.5}M_{\odot}$, is due to a variation in the cold gas accretion of these galaxies that decreases their gas fraction \citep[e.g.,][]{2022A&A...661L...7D} rather than changes in their star formation efficiency.

Our mean molecular gas fraction for MS galaxies are in qualitative and quantitative agreement with the analytical relations of \citet[][]{2019ApJ...887..235L}, \citet[][]{2020ARA&A..58..157T}, and \citet[][]{2022A&A...660A.142W}. Regarding our mean molecular gas depletion time for MS galaxies, they are in agreement with the analytical relations of \citet[][]{2020ARA&A..58..157T} and \citet[][]{2022A&A...660A.142W} at all stellar masses. However, our results only agree with the analytical relation of \citet[][]{2019ApJ...887..235L} for $M_{\star}$ > 10$^{10.5}M_{\odot}$. 
Interestingly, our results on the mean molecular gas fraction agree with those of Liu et al. (2019a) at $M_{\star}$ < 10$^{10.5}M_{\odot}$. The molecular gas depletion time being the ratio between the molecular gas mass and SFR, the discrepancy between our results and the analytical relation of Liu et al. (2019a) at $M_{\star}$ < 10$^{10.5}M_{\odot}$ can only be explained by the use of different SFR$_{\rm MS}$ calibrations. We note that at $z\sim1.4$ and $M_{\star}$ < 10$^{10.5}M_{\odot}$, the results from \citet[][]{2019ApJ...887..235L} are mostly based on extrapolations.

Finally, we find that the slopes of the $\mu_{\rm mol}-\Delta$MS relations might be slightly shallower compared to those found in \citet{2019ApJ...887..235L}, \citet[][]{2020ARA&A..58..157T}, and \citet{2022A&A...660A.142W}; while the slopes of the $\tau_{\rm mol}-\Delta$MS relations appear to be slightly steeper compared to those found in the aforementioned studies. Specifically, while the slopes of the $\mu_{\rm mol}-\Delta$MS and $\tau_{\rm mol}-\Delta$MS relations in logarithm space are $0.51$ and $-0.49$ in \citet[][]{2020ARA&A..58..157T}, we found that slopes of $\sim0.40$ and $\sim-0.65$ would provide a better fit to our measurements, respectively. As SFGs move from the MS to SB, their mean molecular gas fraction increases and the mean molecular gas depletion time decreases slightly more than anticipated by these analytical relations.

\begin{figure}
\centering
\includegraphics[width=\columnwidth]{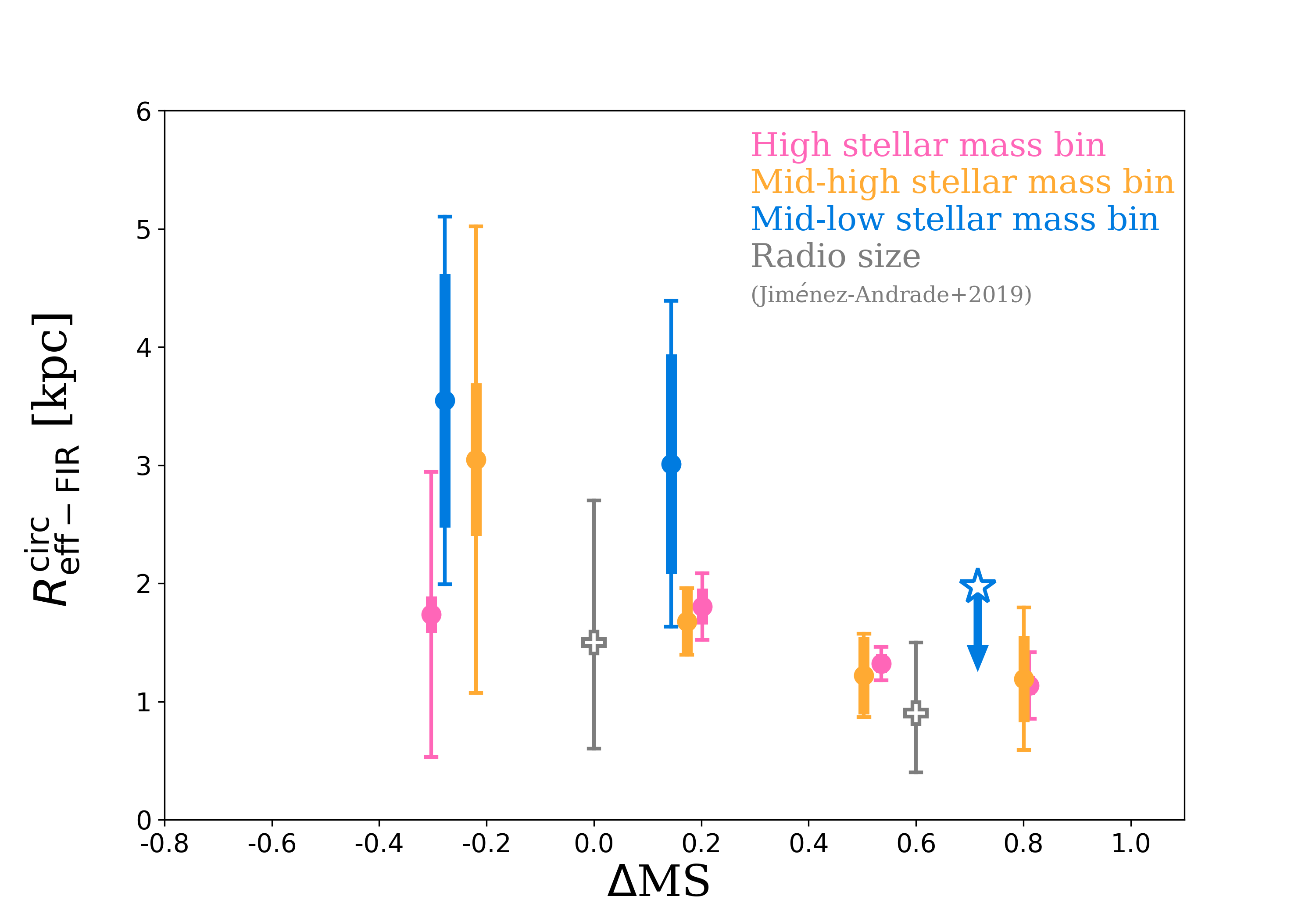}
\caption{Circularized effective FIR radius as a function of distance to the MS. Circles and stars show the mean and upper limit FIR sizes for the resolved and unresolved stacked galaxies, respectively. Error bars are the same as in Fig.~\ref{fig:gas_frac}, but for the circularized effective FIR radius. Crosses are the mean radio sizes of SFGs at $M_{\star}$ > 10$^{10.5}M_{\odot}$ from \citet[][]{2019A&A...625A.114J}.}
\label{fig:FIR_size}
\end{figure}

\subsubsection{Star-forming size of SFGs}
\label{sect:radio_size}
The circularized effective FIR size of our stacked galaxies in different $M_{\star}$--$\Delta$MS bins is shown in Fig.~\ref{fig:FIR_size}. Although still uncertain, these measurements suggest that the FIR size decreases from 2.5~kpc to 1.4~kpc when going from $\Delta$MS$\sim$0 to $\Delta$MS$\sim$0.8, especially when considering only the uncertainties from the instrumental noise (thick error bars in Fig.~\ref{fig:FIR_size}). This finding is in qualitative agreement with \citet[][]{2019A&A...625A.114J}, who studied the radio size of massive (>10$^{10.5}M_{\odot}$) $z\sim$1.5 SFGs as a function of $\Delta$MS. SBs have thus more compact star-forming region compared to MS galaxies, potentially linked to the merger events that triggered them (see Sect.~\ref{sect:size_morphology}). Indeed, simulations suggest that galaxy mergers can trigger the inflow of gas towards the central region and create compact star-forming regions \citep[e.g.,][]{2005MNRAS.361..776S, 2016MNRAS.462.2418S, 2018MNRAS.479.3952B, 2023MNRAS.519.4966B}. 

We also find that the relative dispersion of sizes within the SB population is smaller than within the MS population (see thin error bars in Fig.~\ref{fig:FIR_size}; see also \citet[][]{2019A&A...625A.114J}). 
A small size dispersion within the SB population may seem at odds with the diversity of orientation, size, and mass ratio that a merging system can have and thus the diversity of gas tidal tails or bridges it can exhibit. Our results therefore support a scenario in which the inflow of gas to a central, coalescent, and compact region is short during a merger, i.e., < 0.3~Gyr \citep[e.g.,][]{2022MNRAS.509.2720S}.
On the other hand, the relatively large size dispersion within the MS population, especially for the low $\Delta$MS population, could indicate the presence of a hidden and compact SB population within the normal and extended MS galaxies \citep[see also][]{2018A&A...616A.110E, 2019A&A...625A.114J, 2021MNRAS.508.5217P, 2022A&A...658A..43G}. 

\begin{figure}
\centering
\includegraphics[width=\columnwidth]{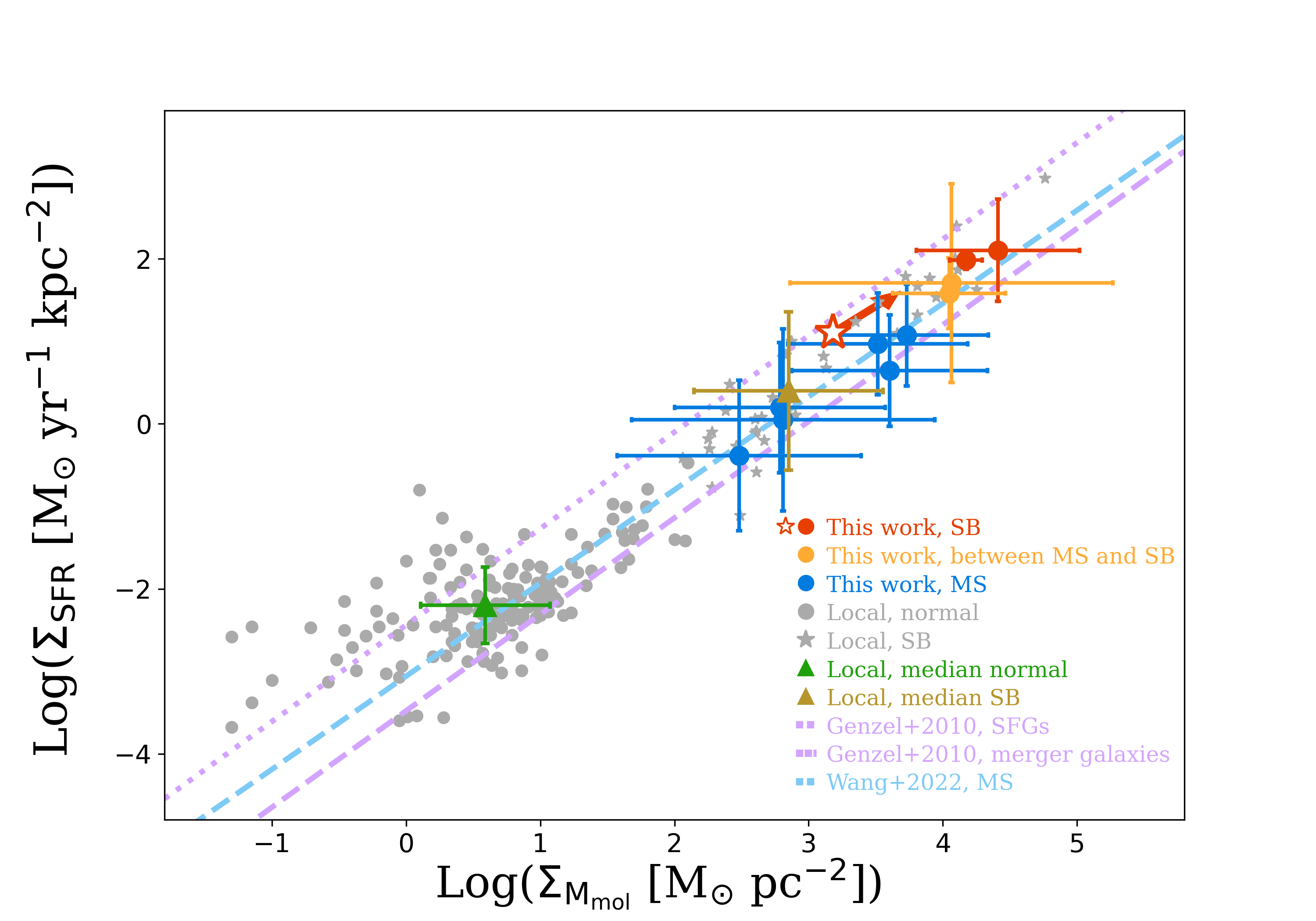}
\caption{Relation between SFR surface density ($\Sigma_{\rm SFR}$) and molecular gas mass surface density ($\Sigma_{\rm M_{mol}}$) of SFGs, i.e., the so-called KS relation. Blue, orange, and red circles represent our stacking results for MS galaxies, in between MS and SB galaxies, and SB galaxies, i.e., $\Delta$MS < 0.4, 0.4$\leq \Delta$MS < 0.7, and 0.7$\leq \Delta$MS, respectively. The red star with the arrow is the lower limit of $\Sigma_{\rm SFR}$ and $\Sigma_{\rm M_{mol}}$ in our highest $\Delta$MS and 10$^{10.0}\leq M_{\star}$/M$_{\odot}$<10$^{10.5}$ bin because their star-forming size is not resolved. Gray circles and stars are local normal and SB galaxies from \citet{1998ApJ...498..541K} and \citet{2019ApJ...872...16D}. Green and brown triangles are the median values of these local normal and SB galaxies. The blue dashed line is the MS-only KS relation from \citet{2022A&A...660A.142W}.
Purple dashed and dotted lines are the KS relations for SFGs and mergers from \citet{2010MNRAS.407.2091G}.}
\label{fig:KS}
\end{figure}

\subsubsection{The Kennicutt-Schmidt relation}
\label{result:KS}
In Fig.~\ref{fig:KS}, we show the relation between the SFR and molecular gas mass surface densities (the so-called KS relation) of our stacked galaxies by combining their SFR, molecular gas mass, and FIR size measurements. We also compare these measurements with results from the literature: molecular gas phase measurement of local normal and starburst galaxies from \citet[][]{1998ApJ...498..541K} and \citet{2019ApJ...872...16D} and the KS relation for SFGs from \citet[][]{2010MNRAS.407.2091G} and \citet[][]{2022A&A...660A.142W}. 

The KS relation of our stacked MS galaxies is fully consistent with the relation for SFGs from \citet[][]{2010MNRAS.407.2091G} and for MS-only galaxies from \citet[][]{2022A&A...660A.142W}. SBs roughly fall on the same KS relation as MS galaxies. Indeed, the increase of the mean star formation efficiency of SBs compared to MS galaxies (see Sect.~\ref{sect:mass_depletion}) is in part explained by the combination of their $\sim$2-3 times more compact FIR sizes (shifting their molecular gas mass surface densities to higher values by a factor $\sim$4-9), $\sim$2.1 times larger molecular gas mass fraction (shifting their molecular gas mass surface densities to higher values by a factor $\sim$2.1), and the slope of 1.1--1.2 of the KS relation.

Our SBs deviate by $\sim$0.5~dex from the KS relation for mergers from \citet[][]{2010MNRAS.407.2091G}. The reason for this difference could be threefold. (i) \citet[][]{2010MNRAS.407.2091G} used CO-based gas mass measurements and an $\alpha_{\rm CO}$ specific for mergers. This particular $\alpha_{\rm CO}$ could underestimate their molecular gas mass by a factor of $\sim$0.5~dex. (ii) \citet[][]{2010MNRAS.407.2091G} averaged the size measurements from H{\footnotesize $\alpha$}, optical, UV, and CO as a representative of star-forming size and report $\sim$2 times larger sizes compared to our FIR measurements. (iii) \citet[][]{2010MNRAS.407.2091G} restricted their SB sample to mergers, while we defined SB based on their distance from the MS.

\begin{figure*}
\centering
\includegraphics[width=1.0\textwidth]{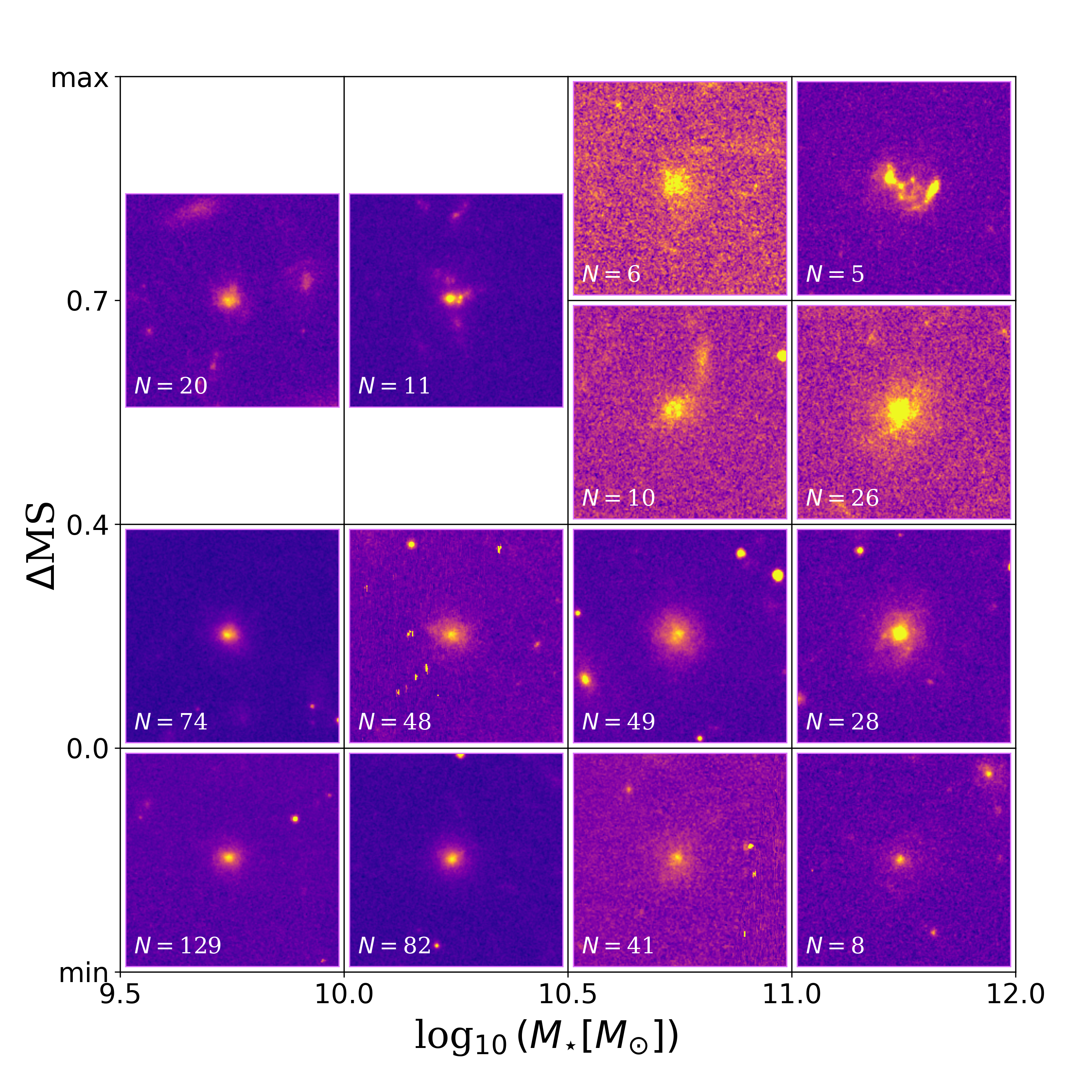}
\caption{Same as Fig.~\ref{fig:ALMA_img}, but for the image stacking on the \textit{HST} $i$-band data.}
\label{fig:HST_img}
\end{figure*}
\begin{figure*}
\centering
\includegraphics[width=1.0\textwidth]{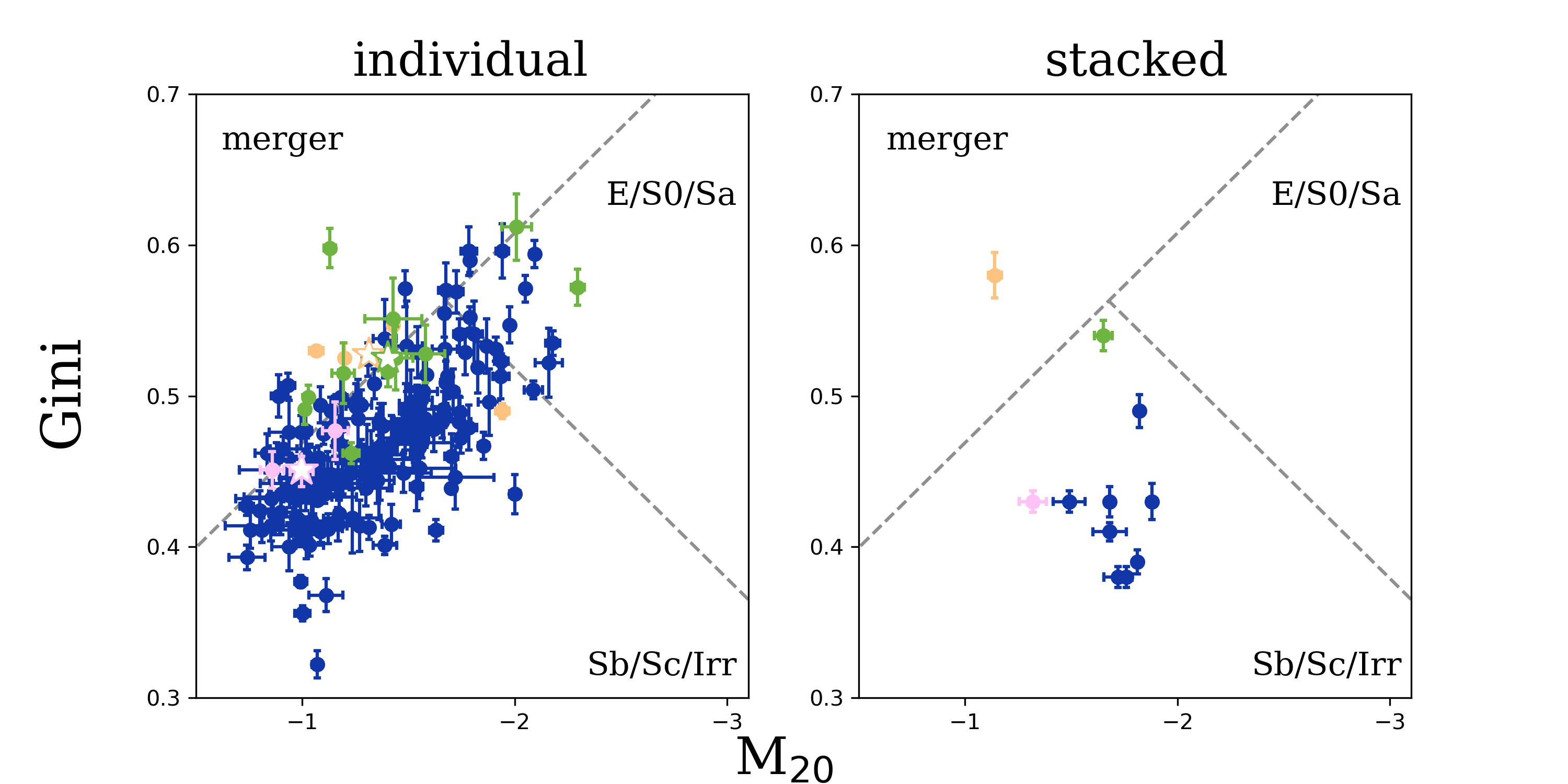}
\caption{Morphological classification of our $M_{\star}$> $10^{10}M_{\odot}$ SFGs measured on their \textit{HST} $i$-band images. Circles in the left panel are the Gini and M20 coefficient measured on \textit{HST} $i$-band images of each individual galaxy, while the right panel shows the Gini and M20 coefficient measured for the stacked images of the different stellar mass and $\Delta$MS bin. Orange, pink, green, and dark blue show SBs ($\Delta$MS > 0.7) at 10$^{11}\leq M_{\star}/{\rm M}_{\odot}<10^{12}$, SBs at 10$^{10.5}\leq M_{\star}/{\rm M}_{\odot}<10^{11}$, SBs at 10$^{10}\leq M_{\star}/{\rm M}_{\odot}<10^{10.5}$, and $M_{\star}$> $10^{10}M_{\odot}$ SFGs without SBs ($\Delta$MS < 0.7), respectively. Stars in the left panel are the median Gini and M20 coefficients of SBs in these three stellar mass bins. Gray dashed lines are the morphology classification from \citet{2008ApJ...672..177L}.}
\label{fig:gini_m20}
\end{figure*}
\begin{figure}
\centering
\includegraphics[width=\columnwidth]{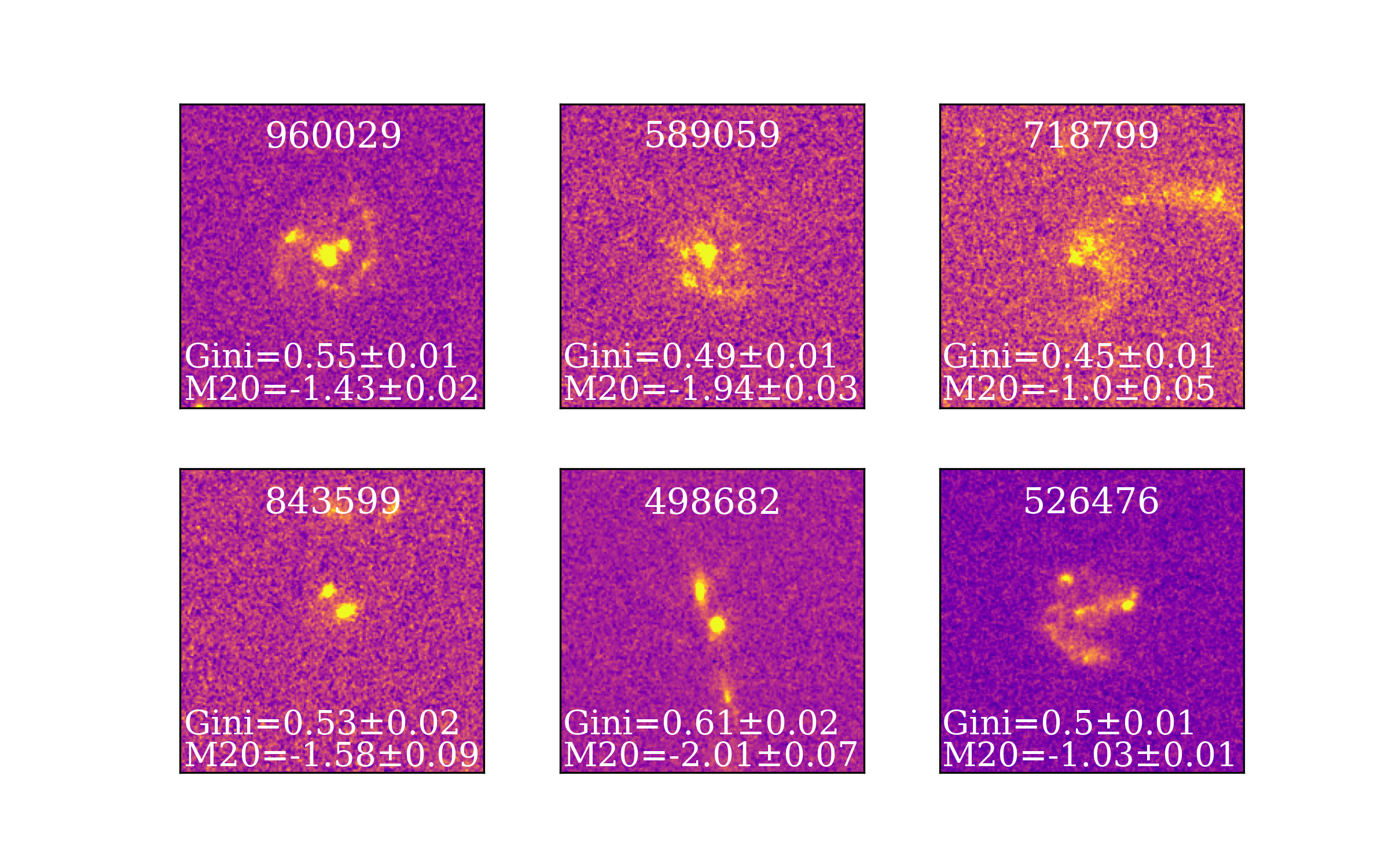}
\caption{Examples of \textit{HST} $i$-band images of individual SB. Numbers on each image show their COSMOS2020 ID and measured Gini and M20 coefficients. Each image has a size of $6\arcsec\times6\arcsec$.}
\label{fig:indi_SB}
\end{figure}
\begin{figure*}
\centering
\includegraphics[width=1.0\textwidth]{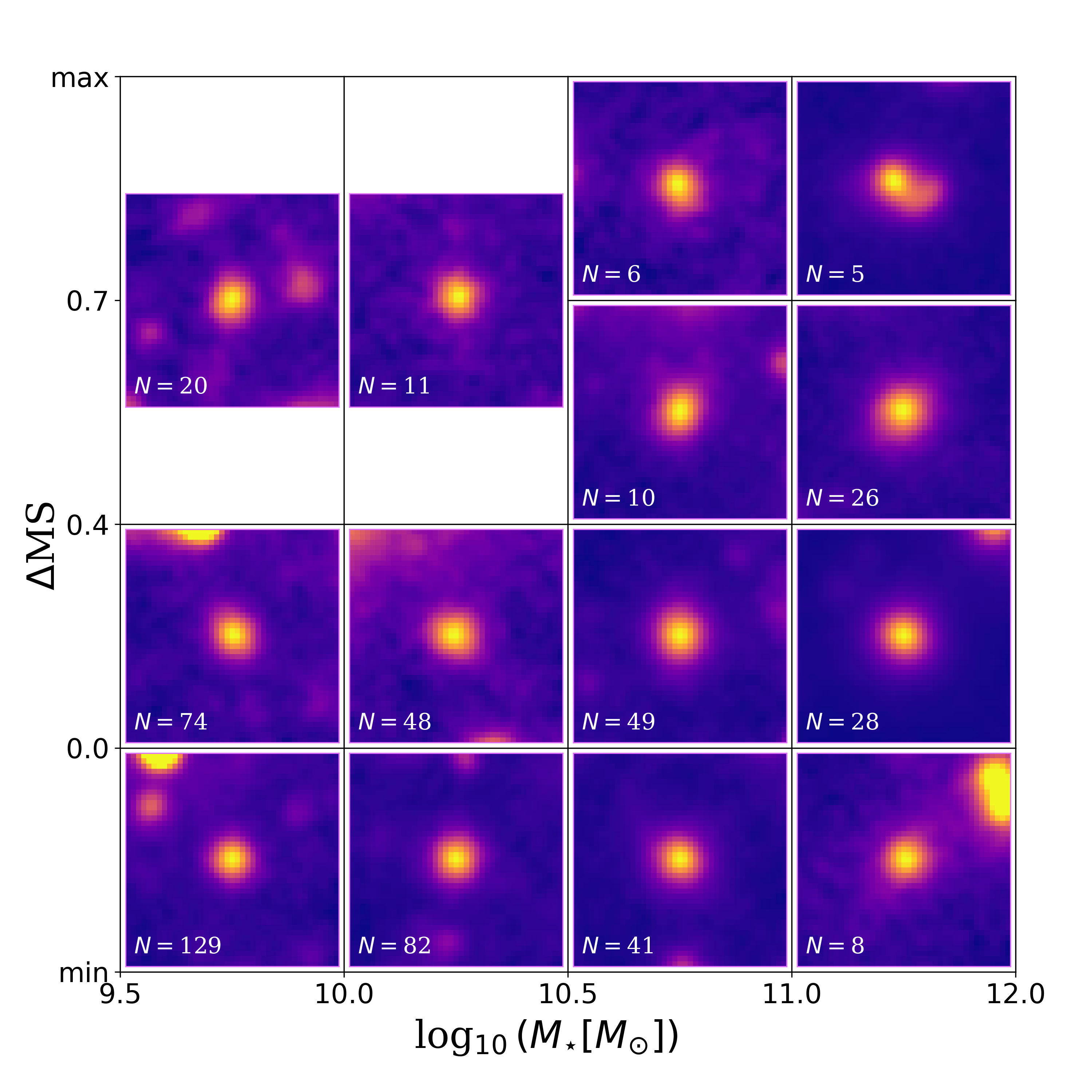}
\caption{Same as Fig.~\ref{fig:ALMA_img}, but for the image stacking on the UltraVISTA $J$-band data.}
\label{fig:VISTA_img}
\end{figure*}
\begin{figure*}
\centering
\includegraphics[width=1.0\textwidth]{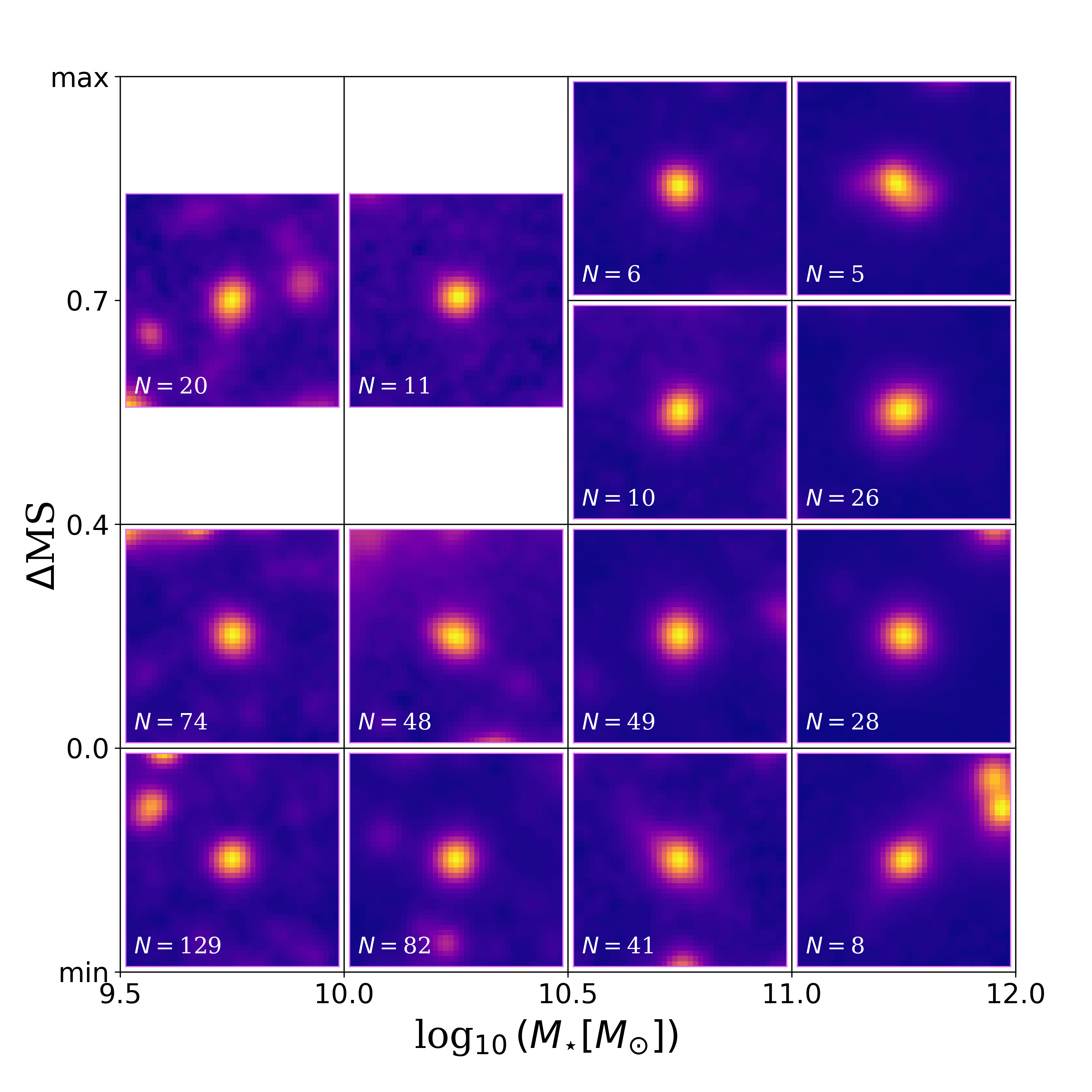}
\caption{Same as Fig.~\ref{fig:ALMA_img}, but for the image stacking on the UltraVISTA $K_{\rm s}$-band data.}
\label{fig:VISTA_K_img}
\end{figure*}
\begin{figure}
\centering
\includegraphics[width=\columnwidth]{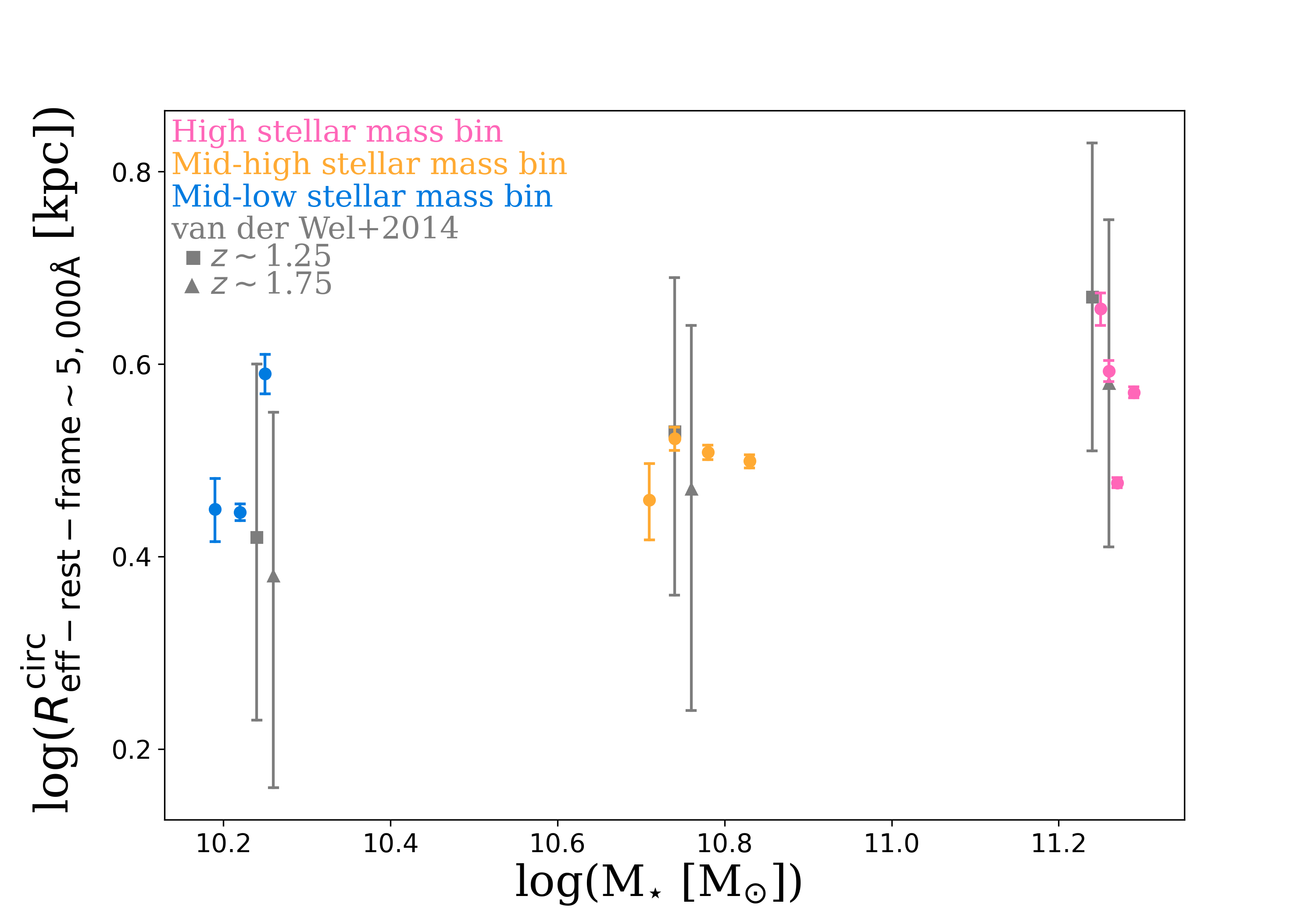}
\caption{Circularized effective radius at rest-frame wavelength of $5,000\AA$ as a function of stellar mass. Circles show the UltraVISTA $J$-band sizes from our stacking analysis, corresponding roughly to a rest-frame wavelength of $5,000\AA$ for our $z\sim1.4$ sample. Squares ($z\sim$1.25) and triangles ($z\sim$1.75), along with their error bars, show the 16th, 50th, and 84th percentiles from \citet{2014ApJ...788...28V} for late-type galaxies as inferred using \textit{HST} images of the CANDELS fields.}
\label{fig:size_mass}
\end{figure}

\subsection{Stellar population size and morphology}
\label{sect:size_morphology}
Results of our image-domain stacking analysis on the \textit{HST} $i$-band dataset are displayed in Fig.~\ref{fig:HST_img}. All the \textit{HST} stacked images have significant detection with a S/N > 5. In addition, while galaxies on the MS ($\Delta$MS < 0.4) exhibit disk-like morphology, it is clear from our stacked \textit{HST} images that galaxies above the MS have disturbed morphology associated with multiple stellar components. To further characterize these disk-like and merger-like morphologies, we measured the Gini and M20 coefficients of the 195 individual galaxies with $M_{\star}$ > 10$^{10}M_{\odot}$ that are sufficiently bright\footnote{Above this stellar mass, 119 galaxies, mostly at low $\Delta$MS, remain undetected in the \textit{HST} $i$ band and are thus excluded from our analysis.} using the python package \texttt{statmorph} \citep[][]{2019MNRAS.483.4140R, 2022ascl.soft01010R}. The Gini describes the relative pixel distribution, while the M20 corresponds to the second-order moment of the brightest 20$\%$ of the pixels of a galaxy \citep[][]{2004AJ....128..163L}. If a galaxy has a merger-like morphology, it has a relatively high Gini and low M20 coefficient compared to disk-like galaxies. The Gini and M20 coefficient measured on the \textit{HST} $i$-band image of each of our galaxies and on their stacked \textit{HST} $i$-band images are shown in Fig.~\ref{fig:gini_m20}. The uncertainties associated with these Gini and M20 coefficient measurements were obtained using Monte Carlo simulations. For each galaxy, we created 100 realizations, each containing the modeled galaxy obtained by \texttt{statmorph} and a random background noise consistent with that found on the residual map of \texttt{statmorph}. We then measured the Gini and M20 coefficients on these 100 realizations and used their distributions as uncertainties. To distinguish merging, spiral, and elliptical galaxies, we used the morphological classification from \citet[][]{2008ApJ...672..177L}. Since we aim to  compare the morphology and size of the stacked galaxies that have at least one FIR size measurement in the stellar mass bin, we did not measure the morphology of galaxies with $M_{\star}$ < $10^{10}M_{\odot}$. 

For individual galaxies (left panel of Fig~\ref{fig:gini_m20}), we find that SBs have indeed a merger-like morphology (see also Fig.~\ref{fig:indi_SB} in which we show examples of these \textit{HST} $i$-band images) For our stacked images (right panel of Fig~\ref{fig:gini_m20}), this separation is less clear as the stacking analysis naturally erases the irregularity of the merging system and thus decreases their Gini and increases their M20 values. Yet stacked SBs are located closer to the merger separation than MS galaxies, and thus still retain some information from their merger origin. We note that this is due to the small number statistic of our SB sample, and we can expect these irregularities to be largely erased in future studies stacking large numbers of SBs.

\begin{figure*}
\centering
\includegraphics[width=1.0\textwidth]{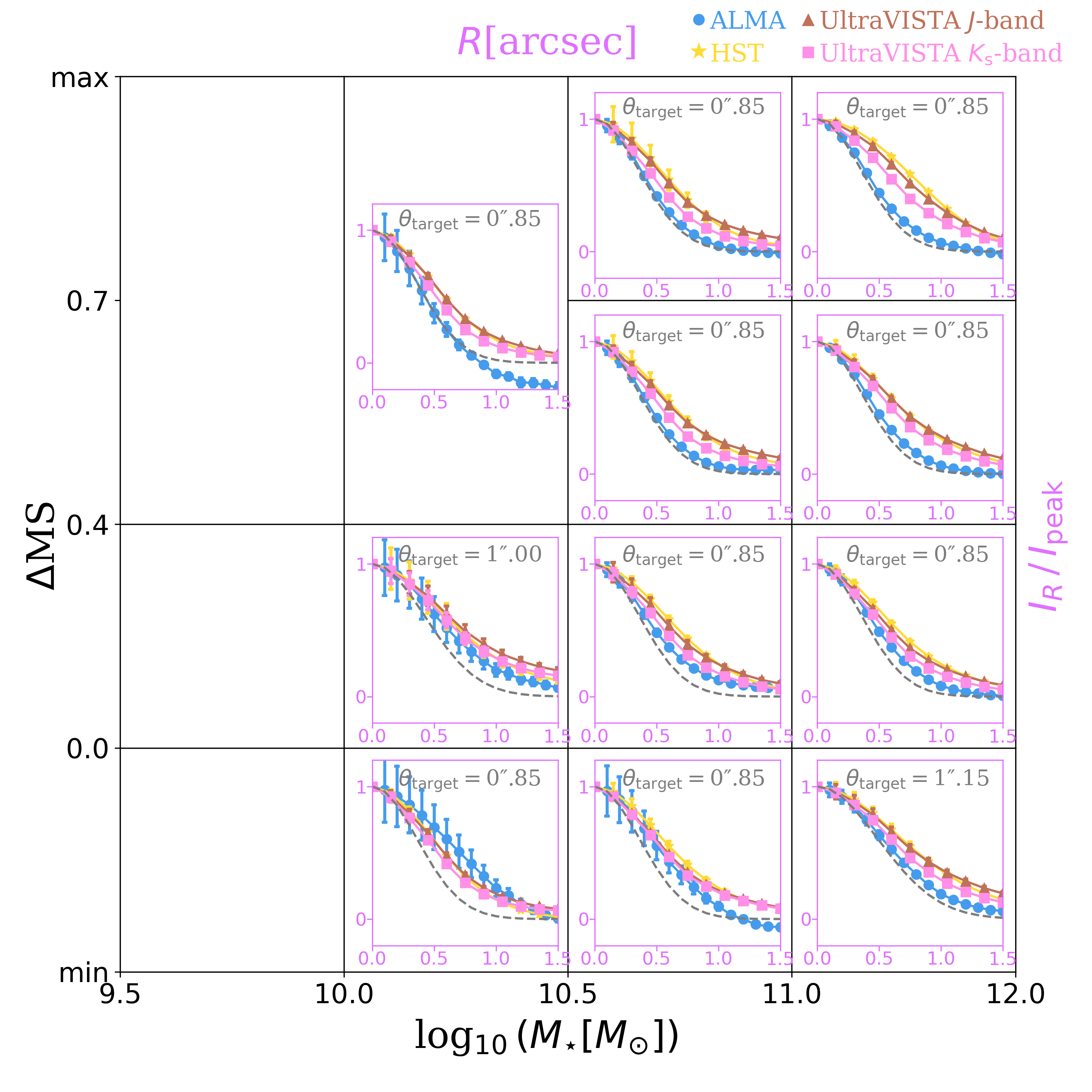}
\caption{Normalized radial light profile of stacked ALMA (blue circles), \textit{HST} (yellow stars), UltraVISTA $J$-band (brown triangles), and UltraVISTA $K_{\rm s}$-band (pink squares) images in different $M_{\star}$--$\Delta$MS bins. All stacked images are matched to the same PSF, i.e., $\theta_{\rm target}$ (gray dashed line). The maximum value of $\Delta$MS for the mid-low, mid-high, and high stellar mass bins are 0.9, 1.0, and 1.1; and the minimum value for the mid-low, mid-high, and high stellar mass bins are -0.7, -0.7, and -0.6, respectively. At the lowest stellar mass, we do not present radial light profiles of the stacked galaxies as they are either unresolved or undetected in our ALMA stacked images.}
\label{fig:light_prof}
\end{figure*}
\begin{figure*}
\centering
\includegraphics[width=1.0\textwidth]{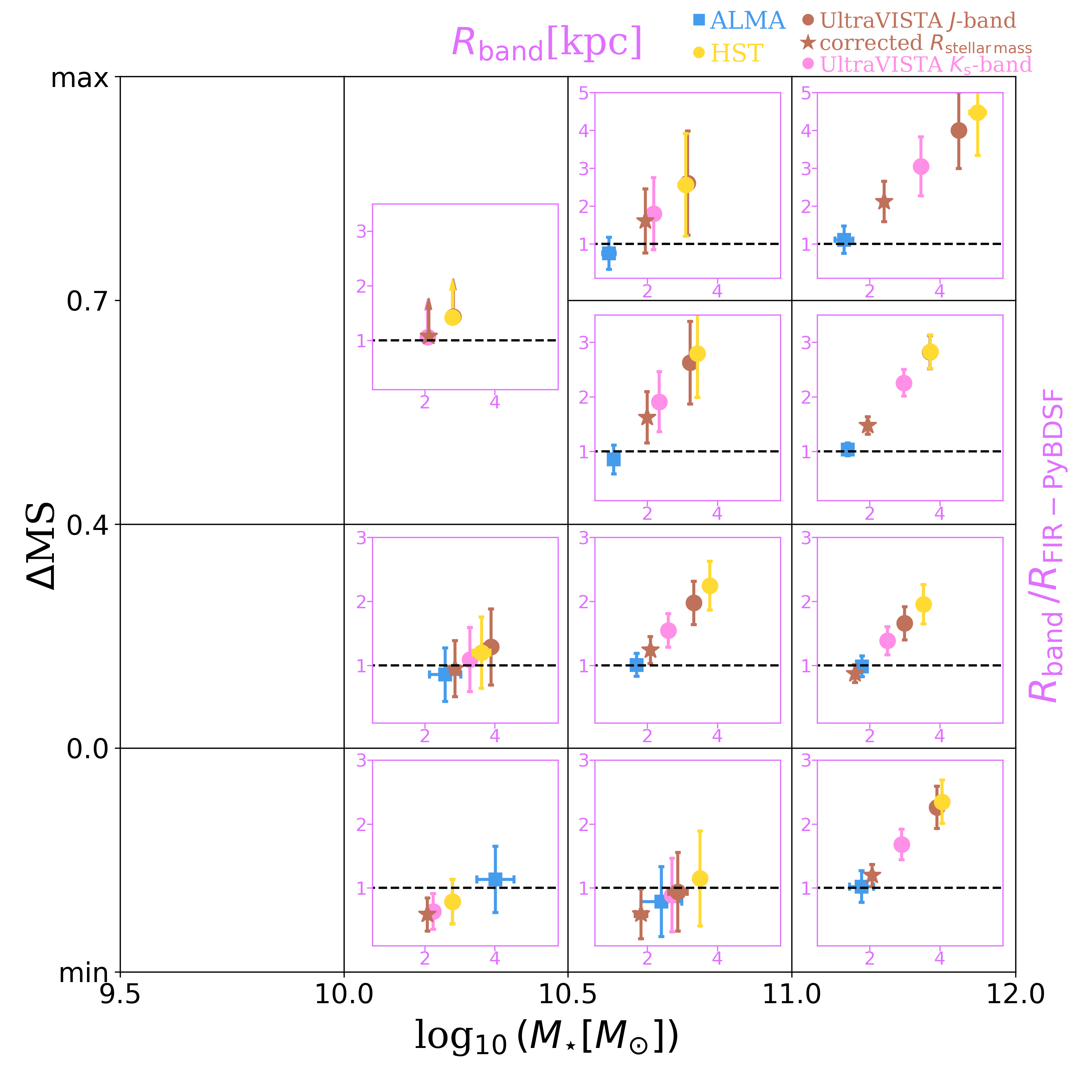}
\caption{Ratio of effective radius in different bands to effective radius in FIR measured by fitting a single Gaussian to our original ALMA stacked images. Yellow, brown, and pink circles correspond the optical size measured from the \textit{HST} $i$-band, UltraVISTA $J$-band, and UltraVISTA $K_{\rm s}$-band light profiles, respectively. Blue squares display the FIR size measured from the ALMA light profile. Brown stars show the stellar mass effective radius inferred by correcting our UltraVISTA $J$-band sizes with a $R_{\rm half-stellar-light}$-to-$R_{\rm half-stellar-mass}$ relation at rest $5000\,\AA$ from \citet[][]{2019ApJ...877..103S}. We provide lower limits of the size ratio in the highest $\Delta$MS and 10$^{10.0}\leq M_{\star}$/M$_{\odot}$<10$^{10.5}$ bin because these galaxies are unresolved in our ALMA stacked images. The maximun value of $\Delta$MS for the mid-low, mid-high, and high stellar mass bin are 0.9, 1.0, and 1.1; and the minimum value for the mid-low, mid-high, and high stellar mass bin are -0.7, -0.7, and -0.6, respectively.}
\label{fig:light_ratio}
\end{figure*}

We show the image-domain stacking analysis for the UltraVISTA dataset in Fig.~\ref{fig:VISTA_img} (UltraVISTA $J$-band) and Fig.~\ref{fig:VISTA_K_img} (UltraVISTA $K_{\rm s}$-band). The stacked images of the UltraVISTA dataset have significant detection with a S/N > 5. As a sanity check, in Fig.~\ref{fig:size_mass} we compare
the mass--size relation at a rest-frame wavelength of $5,000\AA$ inferred from our stacking analysis (i.e., fitting 2D Gaussian profiles to our UltraVISTA $J$-band stacks, our sample being at $z\sim1.4$) to that inferred by \citet{2014ApJ...788...28V} using \textit{HST} images of the CANDELS fields. We find a very good agreement between these two mass--size relations. This demonstrates that, despite the use of a seeing-limited image and of a stacking analysis, we can still obtain accurate estimates of the mean size of SFGs.

This unique FIR, \textit{HST} $i$-band, UltraVISTA $J$-band, and UltraVISTA $K_{\rm s}$-band stacking analysis enables us to compare the size of stacked galaxies at different wavelengths, which in turn reflects the distribution of their various components, such as their young and old stars. Because our stacked FIR and optical images have different spatial resolution, before performing any comparison, we matched the PSFs by convolving each image with different kernels to ensure that all images have the same final resolution. The optical and FIR radial profiles in each $M_{\star}$--$\Delta$MS bin after this PSF-matching are shown in Fig.~\ref{fig:light_prof}. These radial profiles were created by measuring the mean intensity value at different radii and normalizing these mean values with the intensity measured in the central pixel. 

In Fig.~\ref{fig:light_prof}, most of the radial light profiles are more extended than the matched PSF, except for the FIR radial profile in the highest $\Delta$MS and 10$^{10}\leq M_{\star}/{\rm M}_{\odot}<10^{10.5}$ bin. This indicates that except for the latter, all the other bins have spatially resolved emission at optical and FIR wavelengths. However, the spatial extents in these different wavelength are not always the same and we typically found at $M_{\star}$ > 10$^{10.5}M_{\odot}$: \textit{HST} $i$-band > UltraVISTA $J$-band > UltraVISTA $K_{\rm s}$-band > ALMA. To quantify the size ratio between these bands, we measured their deconvolved $FWHM$ (i.e., $\sqrt{FWHM^{2}-\theta_{\rm target}^{2}}$) and converted it to an effective radius following \citet[][]{2019A&A...625A.114J}, i.e., $R_{\rm eff}\approx FWHM\,{\rm (deconvolved)}/2.43$. The effective radius of each band ($R_{\rm band}$) and its ratio to the FIR size is displayed in Fig.~\ref{fig:light_ratio}. We note that the FIR sizes inferred using  this simple beam deconvolution of the light profile are consistent with those inferred using single Gaussian fit to our original (before kernel convolution) stacked ALMA images, with a mean offset of $\sim$5$\%$ (see blue square in Fig.~\ref{fig:light_ratio}). 

We find that the FIR and rest-frame optical sizes of MS galaxies are generally in agreement with each other at $M_{\star}$ <10$^{10.5}M_{\odot}$. Above this stellar mass, their rest-frame optical sizes are usually larger than their FIR sizes. The finding of $R_{\rm optical}$ > $R_{\rm FIR}$ could be interpreted as evidence that we are witnessing the formation of the central bulge in these massive MS galaxies. However, this interpretation depends on which optical sizes we consider, as those tend to decrease when measured at longer wavelengths \citep[\textit{HST} $i$-band > UltraVISTA $J$-band > UltraVISTA $K_{\rm s}$-band; see also, e.g.,][]{2014ApJ...788...28V, 2022ApJ...937L..33S}. It is thus highly probable that our rest-frame optical sizes, even when measured in the longest band available (i.e., UltraVISTA $K_{\rm s}$-band), are still affected by large dust attenuation at the center of these galaxies and could therefore be overestimated. To obtain an accurate comparison between the size of the star-forming component of these galaxies (i.e., their FIR size) and their stellar component, it is thus necessary to determine their half-stellar mass radius from our observed rest-frame optical effective radius. To do so, we corrected our effective radius at rest of $\sim$0.5~$\mu$m (UltraVISTA $J$-band) using the $R_{\rm half-stellar-light}$-to-$R_{\rm half-stellar-mass}$ ratio versus stellar mass relation calibrated at rest of 0.5~$\mu$m by \citet[][]{2019ApJ...877..103S}. The corrected half-stellar mass size of MS galaxies (i.e., those with $\Delta$MS<0.4) are consistent with their stacked FIR size, with  $\langle R_{{\rm stellar}_{(M_{\star},{\rm \Delta MS})}}/R_{{\rm FIR}_{(M_{\star},{\rm \Delta MS})}} \rangle$=0.9$\pm$0.3. This result agrees with the simulation of \citet[][]{2022MNRAS.510.3321P}, which found that the half-stellar mass sizes of SFGs are consistent with the half-light size of the dust continuum emission (and their half SFR size). 

At $\Delta$MS>0.4, using the relation of \citet[][]{2019ApJ...877..103S} to correct half-light size to half-stellar mass size is probably inappropriate as this relation has not been calibrated specifically for SBs. Nevertheless, the fact that at $\Delta$MS>0.4 even the corrected half-stellar mass sizes are larger than the FIR sizes is probably best explained by the fact that these galaxies are dominated by merging systems. In this case, the rest-frame optical sizes are likely dominated by the disrupted morphology of the merging stellar components, whereas the FIR sizes are likely dominated by the coalescing star-forming region as, for example, in the local antennae galaxies \citep[e.g.,][]{2015ApJ...815..103B}.


\section{Discussion}
\label{sec:discussion}
Our analysis reveals several key properties of SFGs at $z\sim$1.4: (i) their mean molecular gas fraction increases and their mean molecular gas depletion time decreases continuously and at similar rates across the MS (-0.2 < $\Delta$MS < 0.2) and into the SB region (i.e., up to $\Delta$MS$\sim$0.8); (ii) their FIR size decreases from 2.5~kpc at $\Delta$MS$\sim$0 to 1.4~kpc at $\Delta$MS$\sim$0.8; (iii) MS galaxies and SBs fall roughly on the same KS relation with a slope of 1.13; (iv) SBs tend to have a merger-like morphology, while MS galaxies tend to have a disk-like morphology; (v) at $M_{\star}$ > 10$^{10.5}M_{\odot}$, their optical sizes are usually $R_{\textit{HST}\,i-{\rm band}}$ > $R_{{\rm UltraVISTA}\,J-{\rm band}}$ > $R_{{\rm UltraVISTA}\,K_{s}-{\rm band}}$, consistently with the so-called negative color gradient; (vi) after having statistically corrected these rest-frame optical sizes for dust attenuation using an empirical relation from \citet{2019ApJ...877..103S}, the stellar mass sizes of MS galaxies are similar to their FIR sizes.

\subsection{The dispersion of the MS}
\label{sect:disp_MS}
At all redshifts and stellar masses, a constant dispersion of $\sim$0.3~dex of the MS has been observed \citep[e.g.,][]{2015A&A...575A..74S, 2020ApJ...899...58L, 2022ApJ...936..165L}. Simulations by \citet[][]{2016MNRAS.457.2790T,2020MNRAS.497..698T} suggest that the mechanisms behind the dispersion of the MS could be complex. For example, gas compaction induced by minor mergers or disk instabilities could enhance the star formation efficiency of galaxies, raise their SFR, and thus increase their $\Delta$MS \citep[e.g.,][]{2012MNRAS.421..818C, 2014MNRAS.438.1870D, 2023MNRAS.tmp.1180L}. On the other hand, the depletion of gas due to star formation or supernova feedback could reduce the gas fraction in galaxies, suppress their SFR, and thus decrease their $\Delta$MS \citep[e.g.,][]{2015MNRAS.446.1939F, 2019ApJ...883...81S}. Most star-forming galaxies would thus experience cycles of the aforementioned mechanisms and oscillate on the MS for a few gigayears, finally quenching into quiescence if the inflow of fresh gas from the the cosmic web cannot support its star formation \citep[e.g.,][]{2015MNRAS.446.1939F, 2021ApJ...908...54W}. During these oscillations within the MS, simulations of \citet[][see also Tacchella et al. \citeyear{2020MNRAS.497..698T}]{2016MNRAS.457.2790T} predict that the variation in molecular gas fraction and star formation efficiency play a similar role. Our findings of both higher molecular gas fraction and lower depletion time as galaxies move across the MS support this scenario. Although still tentative, the observation of slightly more compact FIR sizes at $\Delta$MS$\sim0.2$ than at $\Delta$MS$\sim-0.2$ (see Fig.~\ref{fig:FIR_size}) is also consistent with this scenario.

\subsection{Compact star-forming size or dust attenuation effect}
\label{sect:color_gradient}
To understand the mechanisms leading to the disk growth of SFGs along the MS, one needs to compare their current morphologies (i.e., those of their stellar components) to the distribution of their ongoing star formation. In this context, many studies have investigated the H{\footnotesize $\alpha$} extent ($R_{\rm H{\footnotesize \alpha}}$), i.e., star-forming size, and optical extent ($R_{\rm optical}$), i.e., stellar size, of SFGs at different redshifts and stellar masses \citep[e.g.,][]{2016ApJ...828...27N, 2020ApJ...892....1W, 2022ApJ...937...16M}. They found that $R_{\rm H{\footnotesize \alpha}}$/$R_{\rm optical}$ is usually >1, and that this ratio decreases with decreasing stellar mass. These results suggest that the growth of SFGs follow the so-called inside-out scenario, i.e., where new stars form mainly at large radii. However, recent ALMA observations have contradicted this picture of large star-forming size compared to stellar size at least at the high-mass end ($\geq 10^{10.5}M_{\odot}$) of high-redshift ($z\geq$ 1) SFGs, as those have mostly revealed compact star-forming regions compared to their rest-frame optical extent \citep[e.g.,][]{2018A&A...616A.110E, 2020ApJ...901...74T, 2021MNRAS.508.5217P, 2022A&A...659A.196G, 2022A&A...660A.142W}. This discovery of such compact star-forming regions is usually interpreted as the formation of stellar bulges in SFGs \citep[e.g.,][]{2015MNRAS.450.2327Z, 2016MNRAS.457.2790T}. Recent TNG50 simulations by \citet[][]{2022MNRAS.510.3321P} suggest instead that the observed compact star-forming size in massive SFGs, when compared to their optical sizes, could primarily result from large dust attenuation in the central region. This attenuation strongly affects the estimation of their stellar extent based on rest-frame optical observations. Here we probe for the first time with a mass-complete sample the optical and FIR sizes of MS galaxies down to $M_{\star}$$\sim$10$^{10.0}M_{\odot}$. We found that their star-forming size is similar to their stellar mass size, i.e., $\langle R_{{\rm stellar}_{(M_{\star},{\rm \Delta MS})}}/R_{{\rm FIR}_{(M_{\star},{\rm \Delta MS})}} \rangle$=0.9$\pm$0.3, when accounting for dust attenuation. Our finding supports the interpretation that the observed larger optical size compared to the star-forming size in MS galaxies may be primarily due to dust extinction. 

\subsection{Which mechanism(s) trigger SBs?}
\label{sect:SB_trig}
Starbursts are believed to mostly originate from the major merger of two gas-rich galaxies \citep[e.g.,][]{2019MNRAS.485.1320M, 2022MNRAS.516.4922R, 2023MNRAS.518.3261P}. During a merger, the molecular gas clouds of galaxies can interact with each other and fall into the central region of the merging system \citep[e.g.,][]{2017MNRAS.464.3882W, 2018MNRAS.479.3952B, 2022MNRAS.517.4752S}. The fact that most of the individual SBs have merger-like morphology in the rest-frame optical and the stacked SBs have compact FIR size support such a merger-driven scenario. We note that literature studies also suggest that a portion of SBs could be induced without major merger, such as galaxy interactions or disk instabilities \citep[e.g.,][]{2014MNRAS.444..942P, 2018MNRAS.479..758W}. Our finding that a few individual galaxies do not exhibit merger-like morphologies in their rest-frame optical images (left panel of Fig.~\ref{fig:gini_m20}) supports this interpretation.

While mergers appear to be the dominating driver for the formation of SBs, the exact way galaxy mergers enhance star formation in such systems is debated. Mergers could enhance solely the star formation efficiency \citep[e.g.,][]{2007A&A...468...61D, 2014ApJ...793...19S}. This scenario is observationally supported by the finding in \citet[][]{2010MNRAS.407.2091G} of a distinct KS relation for SBs with ten times higher normalization compared to the KS relation for SFGs. On the other hand, the molecular gas fraction of SFGs could also be elevated by a factor of >2 during a merger, which is revealed by both simulations \citep[e.g.,][]{2010ApJ...720L.149T, 2014MNRAS.442L..33R} and observations in the local Universe \citep[e.g.,][]{2018ApJ...868..132P, 2018MNRAS.476.2591V, 2019A&A...627A.107L}. Indeed, during a merger, the extended H\,{\footnotesize I} reservoir of the merging galaxies \citep[e.g.,][]{2017Sci...355.1285N, 2020ApJ...902..111W} can be compressed and transformed into molecular gas \citep[e.g.,][]{2022ApJ...934..114Y}, thereby enhancing the molecular gas fraction in merging systems. 
We found that the mean molecular gas fraction increases by a factor of $\sim$2.1 from the MS to SBs, while the mean molecular gas depletion time decreases by a factor of $\sim$3.3.
These findings support the scenario where H\,{\footnotesize I} gas from the reservoir flows into the gravitational well of the merging system and subsequently converts into molecular gas during the merger, thereby leading to a higher molecular gas fraction in SBs. 
This mechanism is nearly as important as the variation in star formation efficiency to explain the increase in SFR at a given stellar mass when SFGs move from the MS to the SB regime.

\section{Summary}
\label{sec:summary}
We investigated the molecular gas mass properties of a mass-complete sample (>10$^{9.5}M_{\odot}$) of SFGs at 1.2 $\leq z$ < 1.6, and in particular the variation of these properties with the distance of these galaxies from the main sequence of SFGs (i.e., $\Delta$MS). We applied a $uv$-domain stacking analysis to all archival ALMA data available for the COSMOS field to accurately measure the RJ-based molecular gas mass and size of SFGs in multiple stellar mass and $\Delta$MS bins. With this approach, we studied, for the first time, the molecular gas fraction, molecular gas star formation efficiency, and the KS relation of MS galaxies (-0.7 < $\Delta$MS < 0.4) and SBs ($\Delta$MS > 0.7). Additionally, we performed an image-domain stacking analysis on the \textit{HST} $i$-band and UltraVISTA $J$- and $K_{\rm s}$-band images of these galaxies and measured, thereby, the size and morphology of their stellar component. Our main findings are listed below.
\begin{enumerate}
    \item The mean molecular gas fraction of SFGs increases by a factor of $\sim$2.1, while their mean molecular gas depletion time decreases by a factor of $\sim$3.3, as they move from the MS ($\Delta$MS $\sim$ 0) to SB ($\Delta$MS $\sim$ 0.8).
    \item Across the MS (-0.2 < $\Delta$MS < 0.2), the mean molecular gas fraction of SFGs increases by a factor of $\sim$1.4, while their mean molecular gas depletion time decreases by a factor of $\sim$1.8. 
    \item Although still uncertain, the mean FIR size of SFGs seems to decrease from $\sim$2.5~kpc at the MS to $\sim$1.4~kpc at the SB.
    \item Main-sequence galaxies and SBs follow the same near-linear KS relation, with a slope of $1.1-1.2$.
    \item The measured optical size decreases with wavelength for $M_{\star}$ > 10$^{10.5}M_{\odot}$, with \textit{HST} $i$-band > UltraVISTA $J$-band > UltraVISTA $K_{\rm s}$-band. This is consistent with the so-called negative color gradient described in the literature and is likely due to dust attenuation. 
    \item When accounting for dust attenuation by converting the half-light radius of MS galaxies in the UltraVISTA $J$ band into half-mass radius using the relation of \citet{2019ApJ...877..103S}, the size of the stellar component of MS galaxies is similar to that of their star-forming component.
    \item Starbursts tend to have merger-like morphology, while MS galaxies have disk-like morphology, as revealed by their Gini-M20 coefficient of their optical \textit{HST} $i$-band images.
\end{enumerate}
Overall, the evolution of SFGs across and above the MS is a complex combination of variations in their molecular gas fraction, mean molecular gas depletion time, and mean star-forming size. Variations in both the molecular gas fraction and star formation efficiency seem to drive the oscillation of SFGs within the MS, likely through cycles of compaction, depletion, and outflow, as predicted by \citet[][see also Tacchella et al. \citeyear{2020MNRAS.497..698T}]{2016MNRAS.457.2790T}. SFGs move into the SB region as a result of galaxy mergers that yield to a more compact star-forming region and larger gas fraction that may originate from the transformation of the surrounding H{\footnotesize I} reservoir into molecular gas. This in turn leads to higher $\Sigma_{\rm M_{mol}}$ and thus higher star formation efficiency as SBs follow the universal KS relation with a slope of 1.1--1.2. 

\begin{acknowledgements}
This research was carried out within the Collaborative Research Centre 956, sub-project A1, funded by the Deutsche Forschungsgemeinschaft (DFG) – project ID 184018867. ES acknowledges funding from the European Research Council (ERC) under the European Union's Horizon 2020 research and innovation programme (grant agreement No. 694343). This work was supported by UNAM-PAPIIT IA102023. ALMA is a partnership of ESO (representing its member states), NSF (USA) and NINS (Japan), together with NRC (Canada), MOST and ASIAA (Taiwan), and KASI (Republic of Korea), in cooperation with the Republic of Chile. The Joint ALMA Observatory is operated by ESO, AUI/NRAO and NAOJ. This paper makes use of the following ALMA data: ADS/JAO.ALMA\#2015.1.00055.S, ADS/JAO.ALMA\#2015.1.00137.S, ADS/JAO.ALMA$\#$2015.1.00260.S, ADS/JAO.ALMA$\#$2015.1.00299.S, ADS/JAO.ALMA$\#$2015.1.00379.S, ADS/JAO.ALMA$\#$2015.1.00388.S, ADS/JAO.ALMA$\#$2015.1.00568.S, ADS/JAO.ALMA$\#$2015.1.00664.S, ADS/JAO.ALMA$\#$2015.1.00704.S, ADS/JAO.ALMA$\#$2015.1.01074.S, ADS/JAO.ALMA$\#$2015.1.01105.S, ADS/JAO.ALMA$\#$2015.1.01212.S, ADS/JAO.ALMA$\#$2015.1.01495.S, ADS/JAO.ALMA$\#$2016.1.00279.S, ADS/JAO.ALMA$\#$2016.1.00463.S, ADS/JAO.ALMA$\#$2016.1.00478.S, ADS/JAO.ALMA$\#$2016.1.00624.S, ADS/JAO.ALMA$\#$2016.1.00646.S, ADS/JAO.ALMA$\#$2016.1.00735.S, ADS/JAO.ALMA$\#$2016.1.00778.S, ADS/JAO.ALMA$\#$2016.1.00804.S, ADS/JAO.ALMA$\#$2016.1.01012.S, ADS/JAO.ALMA$\#$2016.1.01040.S, ADS/JAO.ALMA$\#$2016.1.01208.S, ADS/JAO.ALMA$\#$2016.1.01426.S, ADS/JAO.ALMA$\#$2016.1.01454.S, ADS/JAO.ALMA$\#$2016.1.01559.S, ADS/JAO.ALMA$\#$2016.1.01604.S, ADS/JAO.ALMA$\#$2017.1.00326.S, ADS/JAO.ALMA$\#$2017.1.00413.S, ADS/JAO.ALMA$\#$2017.1.00428.L, ADS/JAO.ALMA$\#$2017.1.01176.S, ADS/JAO.ALMA$\#$2017.1.01276.S, ADS/JAO.ALMA$\#$2017.1.01358.S, ADS/JAO.ALMA$\#$2017.1.00046.S, ADS/JAO.ALMA$\#$2017.1.01217.S, ADS/JAO.ALMA$\#$2017.1.01259.S, ADS/JAO.ALMA$\#$2018.1.00085.S, ADS/JAO.ALMA$\#$2018.1.00164.S, ADS/JAO.ALMA$\#$2018.1.00251.S, ADS/JAO.ALMA$\#$2018.1.00635.S, ADS/JAO.ALMA$\#$2018.1.00681.S, ADS/JAO.ALMA$\#$2018.1.00933.S, ADS/JAO.ALMA$\#$2018.1.00992.S, ADS/JAO.ALMA$\#$2018.1.01044.S, ADS/JAO.ALMA$\#$2018.1.01605.S, ADS/JAO.ALMA$\#$2018.1.01871.S, ADS/JAO.ALMA$\#$2019.1.00399.S, ADS/JAO.ALMA$\#$2019.1.00477.S, ADS/JAO.ALMA$\#$2019.1.00964.S, ADS/JAO.ALMA$\#$2019.1.01142.S, ADS/JAO.ALMA$\#$2019.1.01634.L, ADS/JAO.ALMA$\#$2019.1.01702.S.
Based in part on data products produced by TERAPIX and
the Cambridge Astronomy Survey Unit on behalf of the
UltraVISTA consortium.
\end{acknowledgements}

\end{document}